\documentclass[conference]{IEEEtran}

\usepackage{epigraph}
\usepackage{etoolbox}
\usepackage{xspace}
\usepackage[group-separator={,},binary-units]{siunitx}

\usepackage{subcaption}

\usepackage{xspace}
\usepackage[dvipsnames]{xcolor}

\usepackage{float}
\usepackage{graphicx}
\usepackage{amsmath}
\usepackage{footnote}
\usepackage{url}

\usepackage{multirow}
\usepackage{booktabs}
\usepackage{rotating}
\usepackage{pifont}
\usepackage{wasysym}

\usepackage{verbatim}

\newcommand{\ie}{i.\@\,e.,\@\xspace}
\newcommand{\eg}{e.\@\,g.,\@\xspace}

\newcommand{\etc}{etc.\@\xspace}
\newcommand{\cf}{cf.\@\xspace}
\newcommand{\etal}{et~al.\@\xspace}

\ifCLASSINFOpdf
\else
\fi

\begin{document}
%
\title{Adversarial Attacks Against Automatic Speech Recognition Systems via Psychoacoustic Hiding}

\author{\IEEEauthorblockN{Lea Sch\"onherr,
Katharina Kohls,
Steffen Zeiler, 
Thorsten Holz, and
Dorothea Kolossa}
\IEEEauthorblockA{
Horst G\"ortz Institute for IT Security, Ruhr-Universit\"at Bochum,
Germany\\ {\{lea.schoenherr, katharina.kohls, steffen.zeiler, thorsten.holz, dorothea.kolossa\}@rub.de}}
}

\maketitle

\begin{abstract}
Voice interfaces are becoming accepted widely as input methods for a diverse set of devices. This development is driven by rapid improvements in automatic speech recognition~(ASR), which now performs on par with human listening in many tasks. These improvements base on an ongoing evolution of deep neural networks~(DNNs) as the computational core of ASR. However, recent research results show that DNNs are vulnerable to adversarial perturbations, which allow attackers to force the transcription into a malicious output.
   
In this paper, we introduce a new type of adversarial examples based on \emph{psychoacoustic hiding}. Our attack exploits the characteristics of DNN-based ASR systems, where we extend the original analysis procedure by an additional backpropagation step. We use this backpropagation to learn the degrees of freedom for the adversarial perturbation of the input signal, i.e., we apply a psychoacoustic model and manipulate the acoustic signal below the thresholds of human perception. To further minimize the perceptibility of the perturbations, we use forced alignment to find the best fitting temporal alignment between the original audio sample and the malicious target transcription. These extensions allow us to embed an arbitrary audio input with a malicious voice command that is then transcribed by the ASR system, with the audio signal remaining barely distinguishable from the original signal. In an experimental evaluation, we attack the state-of-the-art speech recognition system \emph{Kaldi} and determine the best performing parameter and analysis setup for different types of input. Our results show that we are successful in up to \SI{98}{\percent} of cases with a computational effort of fewer than two minutes for a ten-second audio file. Based on user studies, we found that none of our target transcriptions were audible to human listeners, who still understand the original speech content with unchanged accuracy.
\end{abstract}

\section{Introduction}

\epigraph{Hello darkness, my old friend. I've come to talk with you again. Because a vision softly creeping left its seeds while I was sleeping. And the vision that was planted in my brain still remains, within the sound of silence.}{Simon \& Garfunkel, \emph{The Sound of Silence}}

\textbf{Motivation.}
Deep neural networks~(DNNs) have evolved into the state-of-the-art approach for many machine learning tasks, including automatic speech recognition~(ASR) systems~\cite{xiong2016achieving,saon2017english}. The recent success of DNN-based ASR systems is due to a number of factors, most importantly their power to model large vocabularies and their ability to perform speaker-independent and also highly robust speech recognition. As a result, they can cope with complex, real-world environments that are typical for many speech interaction scenarios such as voice interfaces. In practice,  the importance of DNN-based ASR systems is steadily increasing, \eg within smartphones or stand-alone devices such as Amazon's Echo/Alexa.

On the downside, their success also comes at a price: the number of necessary parameters is significantly larger than that of the previous state-of-the-art Gaussian-Mixture-Model probability densities within Hidden Markov Models (so-called GMM-HMM systems)~\cite{rabiner1993fundamentals}. As a consequence, this high number of parameters gives an adversary much space to explore (and potentially exploit) blind spots that enable her to mislead an ASR system.
Possible attack scenarios include unseen requests to ASR assistant systems, which may reveal private information. Diao \etal demonstrated that such attacks are feasible with the help of a malicious app on a smartphone~\cite{diao2014your}. Attacks over radio or TV, which could affect a large number of victims, are another attack scenario. This could lead to unwanted online shopping orders, which has already happened on normally uttered commands over TV commercials, as Amazon's devices have reacted to the purchase command~\cite{moynihan-17-amazon-dollhouse}. As ASR systems are also often included into smart home setups, this may lead to a significant vulnerability and in a worst-case scenario, an attacker may be able to take over the entire smart home system, including security cameras or alarm systems.

\textbf{Adversarial Examples.}
The general question if ML-based systems can be secure has been investigated in the past~\cite{barreno2006can,barreno2010security, lowd2005adversarial} and some works have helped to elucidate the phenomenon of adversarial examples~\cite{goodfellow2014explaining,fawzi2018analysis,fawzi2016robustness,shaham2015understanding,liu16TAE}. Much recent work on this topic focussed on image classification: different types of adversarial examples have been investigated~\cite{nguyen2015deep,carlini2017towards,evtimov2017robust} and in response, several types of countermeasures have been proposed~\cite{hinton2015distilling,cisse2017parseval,zantedeschi2017efficient}. These countermeasures are focused on only classification-based recognition and some approaches remain resistant~\cite{carlini2017towards}. As the recognition of ASR systems operates differently due to time dependencies, such countermeasures will not work equally in the audio domain. 

In the audio domain, Vaidya \etal were among the first to explore adversarial examples against ASR systems~\cite{vaidya2015cocaine}. They showed how an input signal (\ie audio file) can be modified to fit the target transcription by considering the features instead of the output of the DNN. On the downside, the results show high distortions of the audio signal and a human can easily perceive the attack. Carlini \etal introduced so-called \emph{hidden voice commands} and demonstrated that targeted attacks against HMM-only ASR systems are feasible~\cite{carlini2016hidden}. They use inverse feature extraction to create adversarial audio samples. Still, the resulting audio samples are not intelligible by humans (in most of the cases) and may be considered as noise, but may make thoughtful listeners suspicious. To overcome this limitation, Zhang \etal proposed so-called \emph{DolphinAttacks}: they showed that it is possible to hide a transcription by utilizing non-linearities of microphones to modulate the baseband audio signal with ultrasound higher than 20\,kHz~\cite{zhang2017dolphinattack}. The drawback of this and similar ultrasound-based attacks~\cite{song2017inaudible,roy2017backdoor} is that the attack is costly as the information to manipulate the input features needs to be retrieved from recordings of audio signals with the specific microphone, which is used for the attack. Additionally, the modulation is tailored to a specific microphone, such that the result may differ if another microphone is used. Recently and concurrently, Carlini and Wagner published a technical report in which they introduce a general targeted attack on ASR systems using connectionist temporal classiﬁcation (CTC) loss~\cite{carlini2018audio}. Similarly to previous adversarial attacks on image classifiers, it works with a gradient-descent-based minimization~\cite{carlini2017towards}, but it replaces the loss function by the CTC-loss, which is optimized for time sequences. On the downside, the constraint for the minimization of the difference between original and adversarial sample is also borrowed from adversarial attacks on images and therefore does not consider the limits and sensitivities of human auditory perception. Additionally, the algorithm often does not converge. This is solved by multiple initializations of the algorithm, which leads to high run-time requirements---in the order of hours of computing time---to calculate an adversarial example.
Also very recently, Yuan \etal described \emph{CommanderSong}, which is able to hide transcripts within music~\cite{yuan-18-commandersong}. However, this approach is only shown to be successful in music and it does not contain a human-perception-based noise reduction.

\textbf{Contributions.}
In this paper, we introduce a novel type of adversarial examples against ASR systems based on \emph{psychoacoustic hiding}. We utilize psychoacoustic modeling, as in MP3 encoding, in order to reduce the perceptible noise. For this purpose, hearing thresholds are calculated based on psychoacoustic experiments by Zwicker \etal~\cite{zwicker2013psychoacoustics}. This limits the adversarial perturbations to those parts of the original audio sample, where they are not (or hardly) perceptible by a human. Furthermore, we use backpropagation as one part of the algorithm to find adversarial examples with minimal perturbations. This algorithm has already been successfully used for adversarial examples in other settings~\cite{carlini2017towards,carlini2018audio}. To show the general feasibility of psychoacoustic attacks, we feed the audio signal directly into the recognizer.

A key feature of our approach is the integration of the preprocessing step into the backpropagation. As a result, it is possible to change the raw audio signal without further steps. The preprocessing operates as a feature extraction and is fundamental to the accuracy of an ASR system. Due to the differentiability of each single preprocessing step, we are able to include it in the backpropagation without the necessity to invert the feature extraction. In addition, ASR highly depends on temporal alignment as it is a continuous process. We enhance our attack by computing an optimal alignment with the \emph{forced alignment} algorithm, which calculates the best starting point for the backpropagation. Hence, we make sure to move the target transcription into parts of the original audio sample which are the most promising to not be perceivable by a human. We optimize the algorithm to provide a high success rate and to minimize the perceptible~noise. 

We have implemented the proposed attack to demonstrate the practical feasibility of our approach. We evaluated it against the state-of-the-art DNN-HMM-based ASR system \emph{Kaldi}~\cite{Povey_ASRU2011_2011}, which is one of the most popular toolchains for ASR among researchers~\cite{ali-17-wild, fujimoto-17-cnn,manohar-17-jhu, mcauliffe-17-montreal, ranjan-17-gender, ravanelli-17-distant, trmal-17-kaldi, upadhyaya-17-hindi, villalba-17-autoenc,yuan-18-commandersong} and is also used in commercial products such as Amazon's Echo/Alexa and by IBM and Microsoft~\cite{audhkhasi-17-building, xiong-17-microsoft}. Note that commercial ASR systems do not provide information about their system setup and configuration. 

Such information could be extracted via model stealing and similar attacks (\eg \cite{ilyas2017query,papernot2016transferability,tramer2016stealing,papernot2017blackbox,wang2018hyperparameter}). However, such an end-to-end attack would go beyond the contributions of this work and hence we focus on the general feasibility of adversarial attacks on state-of-the-art ASR systems in a white-box setting. More specifically, we show that it is possible to hide {any target transcription in any audio file with a minimum of perceptible noise in up to \SI{98}{\percent} of cases. We analyze the optimal parameter settings, including different phone rates, allowed deviations from the hearing thresholds, and the number of iterations for the backpropagation. We need less than two minutes on an Intel Core~i7 processor to generate an adversarial example for a ten-second audio file. We also demonstrate that it is possible to limit the perturbations to parts of the original audio files, where they are not (or only barely) perceptible by humans. The experiments show that in comparison to other targeted attacks~\cite{yuan-18-commandersong}, the amount of noise is significantly reduced.

This observation is confirmed during a two-part audibility study, where test listeners transcribe adversarial examples and rate the quality of different settings. The results of the first user study indicate that it is impossible to comprehend the target transcription of adversarial perturbations and only the original transcription is recognized by human listeners. The second part of the listening test is a MUSHRA test~\cite{schinkel2013audio} in order to rate the quality of different algorithm setups. The results show that the psychoacoustic model greatly increases the quality of the adversarial examples. 

\begin{figure*}
    \centering
    \includegraphics[width=1.75\columnwidth]{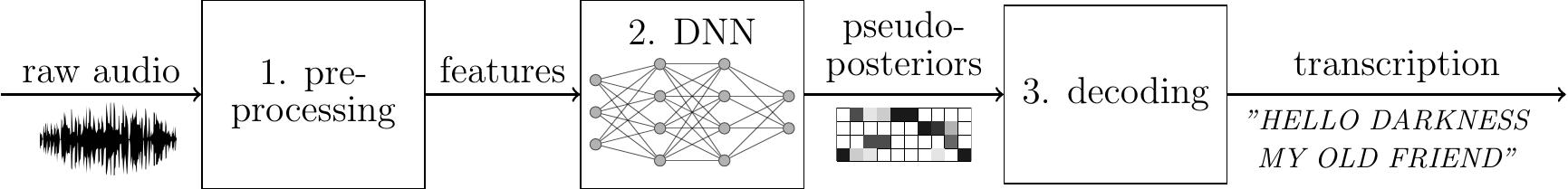}
    \caption{Overview of a state-of-the-art ASR system with the three main components of the ASR system: (1) preprocessing of the raw audio data, (2) calculating pseudo-posteriors with a DNN, and (3) the decoding, which returns the transcription.}
    \label{fig:asr_overview}
\end{figure*}

\smallskip \noindent
In summary, we make the following contributions in this paper:

\begin{itemize}
    \item \textbf{Psychoacoustic Hiding.} We describe a novel type of adversarial examples against DNN-HMM-based ASR systems based on a psychoacoustically designed attack for hiding transcriptions in arbitrary audio files. Besides the psychoacoustic modeling, the algorithm utilizes an optimal temporal alignment and backpropagation up to the raw audio file.
    \item \textbf{Experimental Evaluation.} We evaluate the proposed attack algorithm in different settings in order to find adversarial perturbations that lead to the best recognition result with the least human-perceptible noise.
    \item \textbf{User Study.} To measure the human perception of adversarial audio samples, we performed a user study. More specifically, human listeners were asked to transcribe what they understood when presented with adversarial examples and to compare their overall audio quality compared to original unmodified audio files. 
\end{itemize}

A demonstration of our attack is available online at \url{http://adversarial-asr.selfip.org} where we present several adversarial audio files generated for different kinds of attack scenarios.

\section{Technical Background}

Neural networks have become prevalent in many machine learning tasks, including modern ASR systems.
Formally speaking, they are just functions $y=F(x)$, mapping some input $x$ to its corresponding output $y$.
Training these networks requires the adaptation of hundreds of thousands of free parameters. 
The option to train such models by just presenting input-output pairs during the training process makes deep neural networks (DNNs) so appealing for many researchers.
At the same time, this represents the Achilles' heel of these systems that we are going to exploit for our ASR attack. 
In the following, we provide the technical background as far as it is necessary to understand the details of our approach.

\subsection{Speech Recognition Systems}
\label{sec:asr}

There is a variety of commercial and non-commercial ASR systems available.
In the research community, \emph{Kaldi}~\cite{Povey_ASRU2011_2011} is very popular given that it is an open-source toolkit which provides a wide range of state-of-the-art algorithms for ASR. 
The tool was developed at Johns Hopkins University and is written in C++.
We performed a partial reverse engineering of the firmware of an Amazon Echo and our results indicate that this device also uses \emph{Kaldi} internally to process audio inputs. 
Given \emph{Kaldi}'s popularity and its accessibility, this ASR system hence represents an optimal fit for our experiments. 
Figure~\ref{fig:asr_overview} provides an overview of the main system components that we are going to describe in more detail below.

\subsubsection{Preprocessing Audio Input}\label{sec:prepro}
Preprocessing of the audio input is a synonym for feature extraction: this step transforms the raw input data into features that should ideally preserve all relevant information (\eg phonetic class information, formant structure, etc.), while discarding the unnecessary remainder (\eg properties of the room impulse response, residual noise, or voice properties like pitch information).
For the feature extraction in this paper, we divide the input waveform into overlapping frames of fixed length. 
Each frame is transformed individually using the discrete Fourier transform (DFT) to obtain a frequency domain representation. 
We calculate the logarithm of the magnitude spectrum, a very common feature representation for ASR systems. 
A detailed description is given in Section~\ref{sec:Adv_back}, where we explain the necessary integration of this particular preprocessing into our ASR system.

\subsubsection{Neural Network}
Like many statistical models, an artificial neural network can learn very general input/output mappings from training data. 
For this purpose, so-called neurons are arranged in layers and these layers are stacked on top of each other and are connected by weighted edges to form a DNN. 
Their parameters, \ie~the weights, are adapted during the training of the network.
In the context of ASR, DNNs can be used differently. 
The most attractive and most difficult application would be the direct transformation of the spoken text at the input to a character transcription of the same text at the output. 
This is referred to as an end-to-end-system. 
\emph{Kaldi} takes a different route: it uses a more conventional Hidden Markov Model (HMM) representation in the decoding stage and uses the DNN to model the probability of all HMM states (modeling context-dependent phonetic units) given the acoustic input signal. Therefore, the outputs of the DNN are pseudo-posteriors, which are used during the decoding step in order to find the most likely word sequence.

\subsubsection{Decoding} 
Decoding in ASR systems, in general, utilizes some form of graph search for the inference of the most probable word sequence from the acoustic signal.
In \emph{Kaldi}, a static decoding graph is constructed as a composition of individual transducers (\ie graphs with input/output symbol mappings attached to the edges). 
These individual transducers describe for example the grammar, the lexicon, context dependency of context-dependent phonetic units, and the transition and output probability functions of these phonetic units. The transducers and the pseudo-posteriors (\ie the output of the DNN) are then used to find an optimal path through the word graph.

\subsection{Adversarial Machine Learning}
Adversarial attacks can, in general, be applied to any kind of machine learning system~\cite{barreno2006can, barreno2010security,lowd2005adversarial}, but they are successful especially for DNNs~\cite{papernot2016limitations,goodfellow2014explaining}.

As noted above, a trained DNN maps an input $x$ to an output $y = F(x)$.
In the case of a trained ASR system, this is a mapping of the features into estimated pseudo-posteriors. 
Unfortunately, this mapping is not well defined in all cases due to the high number of parameters in the DNN, which leads to a very complex function~$F(x)$.
Insufficient generalization of~$F(x)$ can lead to blind spots, which may not be obvious to humans. 
We exploit this weakness by using a manipulated input $x'$ that closely resembles the original input $x$, but leads to a different mapping:
\begin{gather*}
x' = x + \delta, \quad \textrm{such that } F(x) \neq F(x'),
\end{gather*}
where we minimize any additional noise~$\delta$ such that it stays close to the hearing threshold. For the minimization, we use a model of human audio signal perception. This is easy for cases where no specific target~$y'$ is defined. In the following, we show that adversarial examples can even be created very reliably for targeted attacks, where the output~$y'$ is defined.

\subsection{Backpropagation} 
Backpropagation is an optimization algorithm for computational graphs (like those of neural networks) based on gradient descent. It is normally used during the training of DNNs to learn the optimal weights. With only minor changes, it is possible to use the same algorithm to create adversarial examples from arbitrary inputs. For this purpose, the parameters of the DNN are kept unchanged and only the input vector is updated.

For backpropagation, three components are necessary:
\begin{enumerate}
    \item \textbf{Measure loss.} The difference between the actual output~$y_i = F(x_i)$ and the target output~$y'$ is measured with a loss function~$L(y_i,y')$. The index~$i$ denotes the current iteration step, as backpropagation is an iterative algorithm. The cross-entropy, a commonly used loss function for DNNs with classification tasks, is employed  here
    \begin{equation*}
    L(y_i,y') =  - \sum y_i \log(y'). 
\end{equation*}   
    \item \textbf{Calculate gradient.} The loss is back-propagated to the input~$x_i$ of the neural network. For this purpose, the gradient~$\nabla x_i$ is calculated by partial derivatives and the chain rule
    \begin{equation}
\nabla x_i = \frac{\partial L(y_i,y')}{\partial x_i} = \frac{\partial L(y_i,y')}{\partial F(x_i)} \cdot \frac{\partial  F(x_i)}{\partial x_i}.
\label{eq:gradient}
\end{equation}

The derivative of $F(x_i)$ depends on the topology of the neural network and is also calculated via the chain rule, going backward through the different layers.
 
    \item \textbf{Update.} The input is updated according to the back-propagated gradient and a learning rate~$\alpha$ via
    \begin{equation*}
    x_{i+1} = x_i - \nabla x_i  \cdot \alpha.
\end{equation*}   
\end{enumerate} 

These steps are repeated until convergence or until an upper limit for the number of iterations is reached. 
With this algorithm, it is possible to approximately solve problems iteratively, which cannot be solved analytically. Backpropagation is guaranteed to find a minimum, but not necessarily the global minimum. As there is not only one solution for a specific target transcription, it is sufficient for us to find \emph{any} solution for a valid adversarial example.

\subsection{Psychoacoustic Modeling}
\label{sec:psychoac}
Psychoacoustic hearing thresholds describe how the dependencies between frequencies lead to masking effects in the human perception. Probably the best-known example for this is MP3 compression~\cite{ISO11172}, where the compression algorithm applies a set of empirical hearing thresholds to the input signal. By removing those parts of the input signal that are inaudible by human perception, the original input signal can be transformed into a smaller but lossy representation.

\subsubsection{Hearing Thresholds}
MP3 compression depends on an empirical set of hearing thresholds that define how dependencies between certain frequencies can mask, \ie make inaudible, other parts of an audio signal. When applied to the frequency domain representation of an input signal, the thresholds indicate which parts of the signal can be altered in the following quantization step, and hence, help to compress the input. We utilize this psychoacoustic model for our manipulations of the signal, \ie we apply it as a rule set to \emph{add} inaudible noise. We derive the respective set of thresholds for an audio input from the psychoacoustic model of MP3 compression. In Figure~\ref{fig:masking_example} an example for a single tone masker is shown. Here, the green line represents the human hearing thresholds in quiet over the complete human-perceptible frequency range. In case of a masking tone, this threshold increases, reflecting the decrease in sensitivity in the frequencies around the test tone. In Figure~\ref{fig:masking_example} this is shown for \SI{1}{kHz} and \SI{60}{dB}.

\begin{figure}
    \centering
    \includegraphics[width=.8\columnwidth]{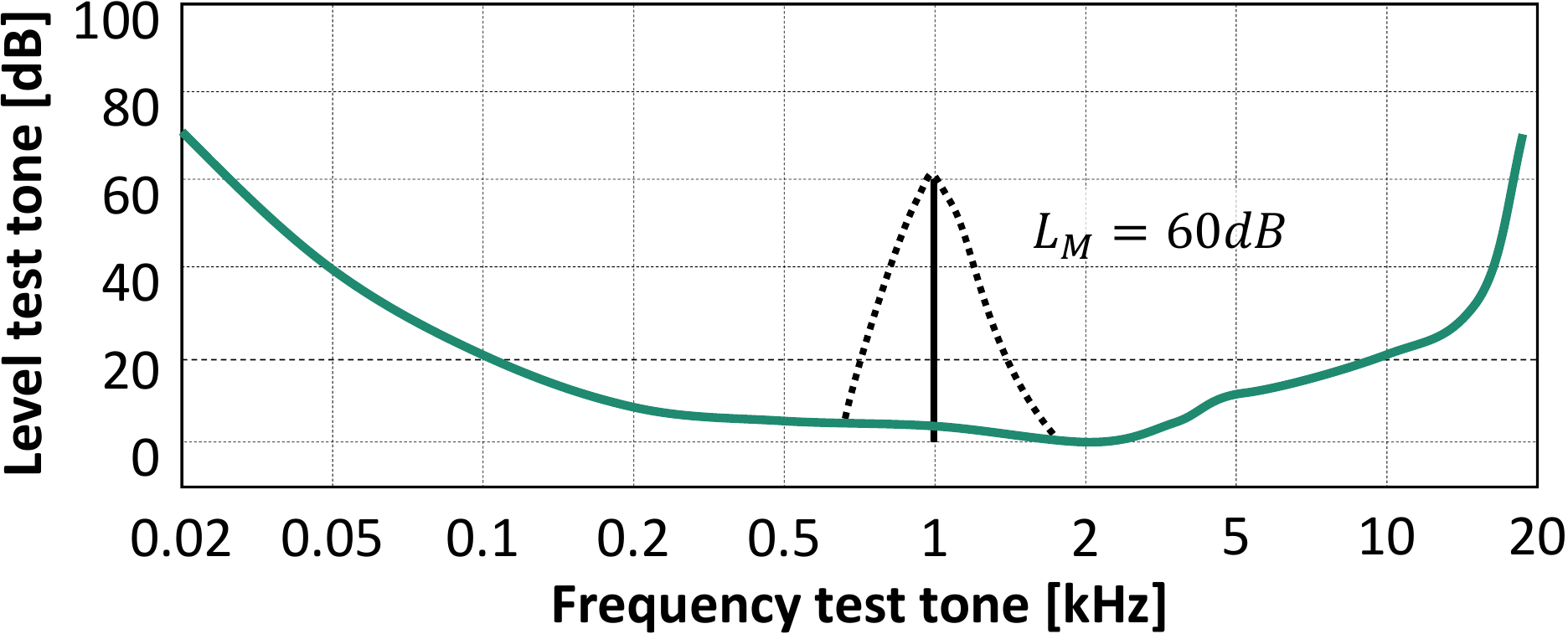}
    \caption{Hearing threshold of test tone (dashed line) masked by a $L_{CB}=60dB$ tone at \SI{1}\,kHz~\cite{zwicker2013psychoacoustics}. In green, the hearing threshold in quiet is shown.}
    \label{fig:masking_example}
\end{figure}

\subsubsection{MP3 Compression}
We receive the original input data in buffers of \num{1024} samples length that consist of two \num{576} sample \emph{granule} windows. One of these windows is the current granule, the other is the previous granule that we use for comparison. We use the fast Fourier transform to derive \num{32} frequency bands from both granules and break this spectrum into MPEG~ISO~\cite{ISO11172} specified scale factor bands. This segmentation of frequency bands helps to analyze the input signal according to its acoustic characteristics, as the hearing thresholds and masking effects directly relate to the individual bands. We measure this segmentation of bands in \emph{bark}, a subjective measurement of frequency. Using this bark scale, we estimate the \emph{relevance} of each band and compute its energy.

In the following steps of the MP3 compression, the thresholds for each band indicate which parts of the frequency domain can be removed while maintaining a certain audio quality during quantization. In the context of our work, we use the hearing thresholds as a guideline for acceptable manipulations of the input signal. They describe the amount of energy that can be added to the input in each individual window of the signal. An example of such a matrix is visualized in Figure~\ref{fig:spectro_thresholds}. The matrices are always normalized in such a way that the largest time-frequency-bin energy is limited to \SI{95}{\dB}.

\section{Attacking ASR via Psychoacoustic Hiding}

In the following, we show how the audible noise can be limited by applying hearing thresholds during the creation of adversarial examples. 
As an additional challenge, we need to find the optimal temporal alignment, which gives us the best starting point for the insertion of malicious perturbations.
Note that our attack integrates well into the DNN-based speech recognition process: we use the trained ASR system and apply backpropagation to update the input, eventually resulting in adversarial examples. A demonstration of our attack is available at \url{http://adversarial-asr.selfip.org}.

\subsection{Adversary Model}

Throughout the rest of this paper, we assume the following adversary model.
First, we assume a white-box attack, where the adversary knows the ASR mechanism of the attacked system. Using this knowledge, the attacker generates audio samples containing malicious perturbations before the actual attack takes place, \ie the attacker exploits the ASR system to obtain an audio file that produces the desired recognition result.
Second, we assume the ASR system to be configured in such a way that it gives the best possible recognition rate. In addition, the trained ASR system, including the DNN, remains unchanged over time. 
Finally, we assume a perfect transmission channel for replaying the manipulated audio samples, hence, we do not take perturbations through audio codecs, compression, hardware, \etc into account by feeding the audio file directly into the recognizer.
Note that we only consider targeted attacks, where the target transcription is predefined (\ie the adversary chooses the target sentence).

\label{sec:attack}
\begin{figure}
    \centering
    \includegraphics[width=\columnwidth]{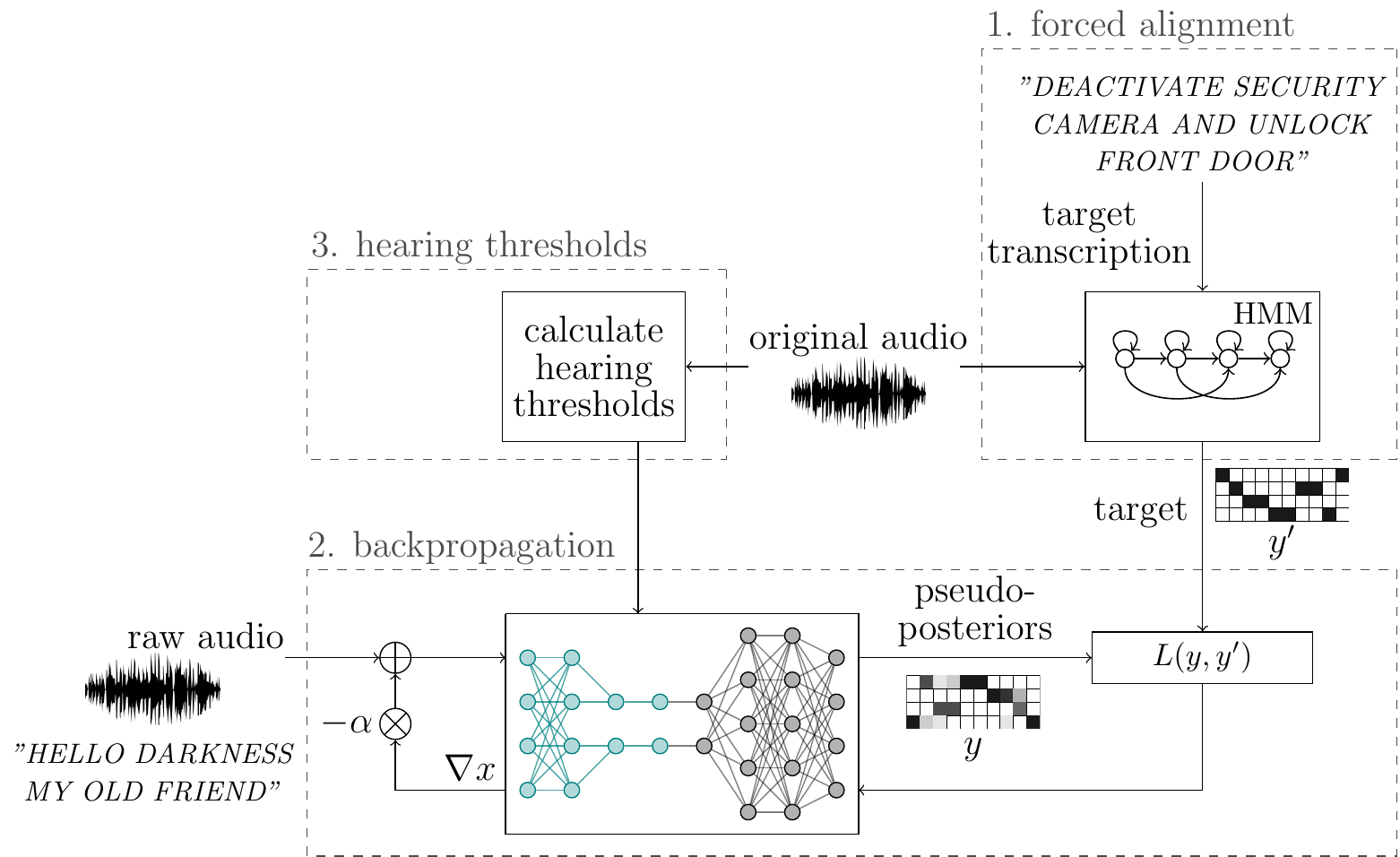}
    \caption{The creation of adversarial examples can be divided into three components: (1) \emph{forced alignment} to find an optimal target for the (2) backpropagation and the integration of (3)~the hearing thresholds.}
    \label{fig:adv+dnn}
\end{figure}

\subsection{High-Level Overview}

The algorithm for the calculation of adversarial examples can be divided into three parts, which are sketched in Figure~\ref{fig:adv+dnn}.
Before the backpropagation, the best possible temporal alignment is calculated via so-called \emph{forced alignment}. The algorithm uses the original audio signal and the target transcription as inputs in order to find the best target pseudo-posteriors. The \emph{forced alignment} is performed once at the beginning of the algorithm.

With the resulting target, we are able to apply backpropagation to manipulate our input signal in such a way that the speech recognition system transcribes the desired output. The backpropagation is an iterative process and will, therefore, be repeated until it converges or a fixed upper limit for the number of iterations is reached. 

The hearing thresholds are applied during the backpropagation in order to limit the changes that are perceptible by a human. The hearing thresholds are also calculated once and stored for the backpropagation. A detailed description of the integration is provided in Section~\ref{sec:hearing_thresholds}.

\subsection{Forced Alignment}

One major problem of attacks against ASR systems is that they require the recognition to pass through a certain sequence of HMM states in such a way that it leads to the target transcription. However, due to the decoding step---which includes a graph search---for a given transcription, many valid pseudo-posterior combinations exist. For example, when the same text is spoken at different speeds, the sequence of the HMM states is correspondingly faster or slower. We can benefit from this fact by using that version of pseudo-posteriors which best fits the given audio signal and the desired target transcription.

We use forced alignment as an algorithm for finding the best possible temporal alignment between the acoustic signal that we manipulate and the transcription that we wish to obtain. This algorithm is provided by the \emph{Kaldi} toolkit. 
Note that it is not always possible to find an alignment that fits an audio file to any target transcription. In this case, we set the alignment by dividing the audio sample equally into the number of states and set the target according to this division.

\subsection{Integrating Preprocessing}
We integrate the preprocessing step and the DNN step into one joint DNN. This approach is sketched in Figure~\ref{fig:pre+dnn}. The input for the preprocessing is the same as in Figure~\ref{fig:asr_overview}, and the pseudo-posteriors are also unchanged. 

This design choice does not affect the accuracy of the ASR system, but it allows for manipulating the raw audio data by applying backpropagation to the preprocessing steps, directly giving us the optimally adversarial audio signal as result. 

\subsection{Backpropagation}
\label{sec:Adv_back}

Due to this integration of preprocessing into the DNN, Equation~\eqref{eq:gradient} has to be extended to
 \begin{equation*}
\nabla x = \frac{\partial L(y,y')}{\partial F(\chi)} \cdot \frac{\partial F(\chi)}{\partial  F_{P}(x)} \cdot \frac{\partial F_{P}(x)}{\partial x},
\label{eq:ext_gradient}
\end{equation*}

where we ignore the iteration index $i$ for simplicity.
All preprocessing steps are included in $\chi = F_P(x)$ and return the input features~$\chi$ for the DNN.
In order to calculate $\frac{\partial F_{P}(x)}{\partial x}$, it is necessary to know the derivatives of each of the four preprocessing steps.
We will introduce these preprocessing steps and the corresponding derivatives in the following. 

\subsubsection{Framing and Window Function}

In the first step, the raw audio data is divided into $T$ frames of length~$N$ and a window function is applied to each frame. A window function is a simple, element-wise multiplication with fixed values~$w(n)$
\begin{equation*}
x_w(t,n) = x(t,n) \cdot w(n), \quad n = 0, \dots, N-1,
\end{equation*}
with $t = 0, \dots, T-1$. Thus, the derivative is just
\begin{equation*}
\frac{\partial x_w(t,n)}{\partial x(t,n)} = w(n).
\end{equation*}

\subsubsection{Discrete Fourier Transform}
For transforming the audio signal into the frequency domain, we apply a DFT to each frame $x_w$. This transformation is a common choice for audio features. The DFT is defined as 
\begin{equation*}
X(t,k) = \sum^{N-1}_{n=0}x_w(t,n)e^{-i2\pi \frac{kn}{N}}, \quad k = 0, \dots, N-1.
\end{equation*}

In Figure~\ref{fig:pre+dnn} the DFT layer is shown schematically as Layer~$2$ of the preprocessing sub-DNN. Since the DFT is a weighted sum with fixed coefficients~$e^{-i2\pi \frac{kn}{N}}$, the derivative for the backpropagation is simply the corresponding coefficient
\begin{equation*}
\frac{\partial X(t,k)}{\partial x_w(t,n)} = e^{-i2\pi \frac{kn}{N}}, \quad k,n = 0, \dots, N-1.
\end{equation*}
\subsubsection{Magnitude}
The output of the DFT is complex valued, but as the phase is not relevant for speech recognition, we just use the magnitude of the spectrum, which is defined as 
\begin{align*}
|X(t,k)|^2 &=  a(t,k)^2 + b(t,k)^2, \\
\textrm{with} \quad  a(t,k) &=  \operatorname{Re}(X(t,k)), \\
 \quad b(t,k) & =  \operatorname{Im}(X(t,k)),
\label{eq:magintude}
\end{align*}
with $\operatorname{Re}(X(t,k))$ and $\operatorname{Im}(X(t,k))$ as the real and imaginary part of $X(t,k)$.
For the backpropagation, we need the derivative of the magnitude. In general, this is not well defined and allows two solutions,
\begin{equation*}
   \frac{\partial |X(t,k)|^2}{\partial X(t,k)} =
   \begin{cases}
     2 \cdot a(t,k)  \\
     2 \cdot b(t,k) 
   \end{cases}.
\end{equation*}
We circumvent this problem by considering the real and imaginary parts separately and calculate the derivatives for both cases
\begin{equation}
\nabla X(t,k) = \begin{pmatrix} 
\frac{\partial |X(t,k)|^2}{\partial \operatorname{Re}(X(t,k))} \\
\frac{\partial |X(t,k)|^2}{\partial \operatorname{Im}(X(t,k))} 
\end{pmatrix} = \begin{pmatrix} 
2 \cdot \operatorname{Re}(X(t,k)) \\
2 \cdot \operatorname{Im}(X(t,k)) 
\end{pmatrix}.
   \label{eq:grad_dft}
\end{equation}
This is possible, as real and imaginary parts are stored separately during the calculation of the DNN, which is also sketched in Figure~\ref{fig:pre+dnn}, where pairs of nodes from layer~$2$ are connected with only one corresponding node in layer~$3$. Layer~$3$ represents the calculation of the magnitude and therefore halves the data size.

\subsubsection{Logarithm}
The last step is to form the logarithm of the squared magnitude $\chi = \text{log}(|X(t,k)|^2)$, which is the common feature representation in speech recognition systems. It is easy to find its derivative as
\begin{equation*}
   \frac{\partial \chi}{\partial |X(t,k)|^2} = \dfrac{1}{|X(t,k)|^2}.
\end{equation*}
\begin{figure}
    \centering
    \includegraphics[width=\columnwidth]{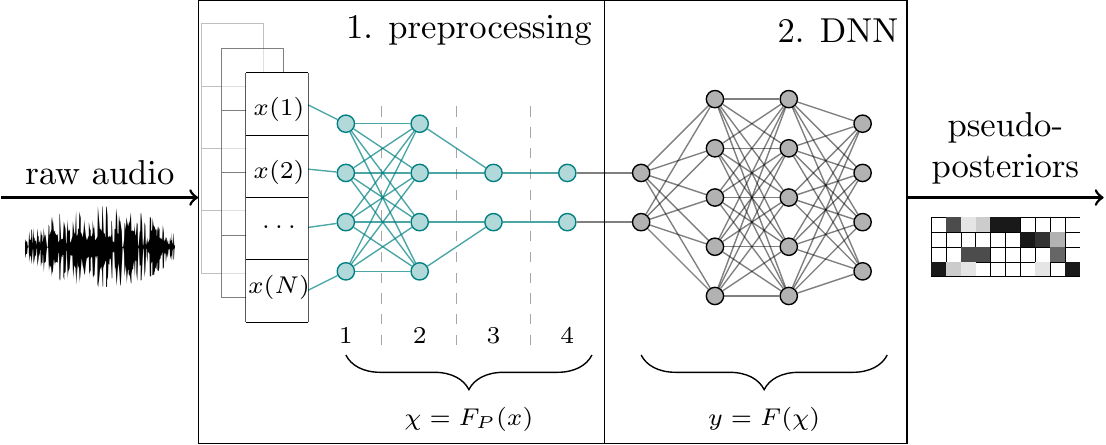}
    \caption{For the creation of adversarial samples, we use an ASR system where the preprocessing is integrated into the~DNN. Layers 1--4 represent the separate preprocessing~steps.}
    \label{fig:pre+dnn}
\end{figure}

\begin{figure*}[!h]
  \centering
  
  \begin{subfigure}[t]{0.45\textwidth}
  \centering
  \includegraphics[width=\textwidth]{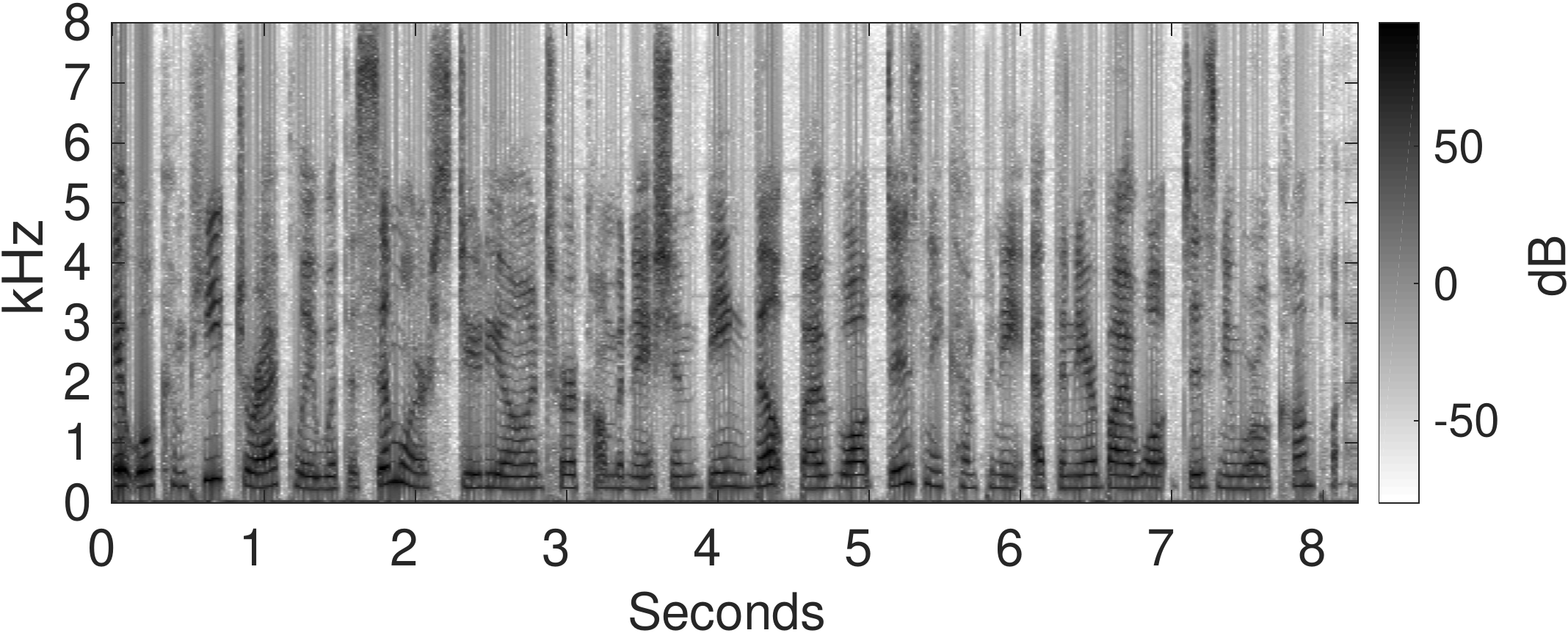}
  \caption{Original audio signal power spectrum~$|\mathbf S|$ with transcription: \emph{\scriptsize{``THE DISNEY PROJECT IS SCHEDULED FOR COMPLETION IN NINETEEN EIGHTY EIGHT AT AN ESTIMATED COST OF TWO HUNDRED AND FIFTY~MILLION.''}}}
  \label{fig:spectro_original}
  \end{subfigure}%
  \hfill
  \begin{subfigure}[t]{0.45\textwidth}
  \centering
  \includegraphics[width=\textwidth]{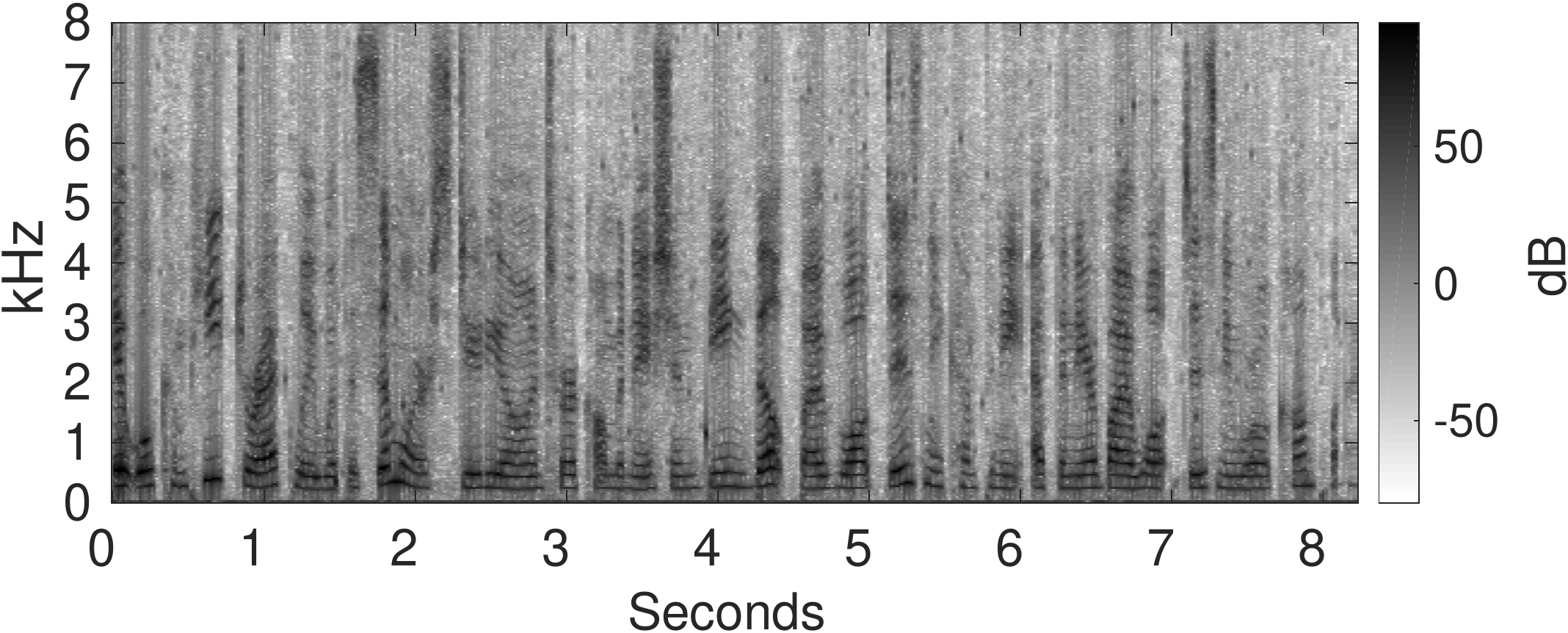}
  \caption{Adversarial audio signal power spectrum~$|\mathbf M|$ with transcription:  \emph{\scriptsize{``I AM A SPACE INVADER COMING FOR YOU.''}}}
  \label{fig:spectro_spoofed}
  \end{subfigure}
  
  \bigskip 
  
  \begin{subfigure}[t]{0.45\textwidth}
  \centering
  \includegraphics[width=\textwidth]{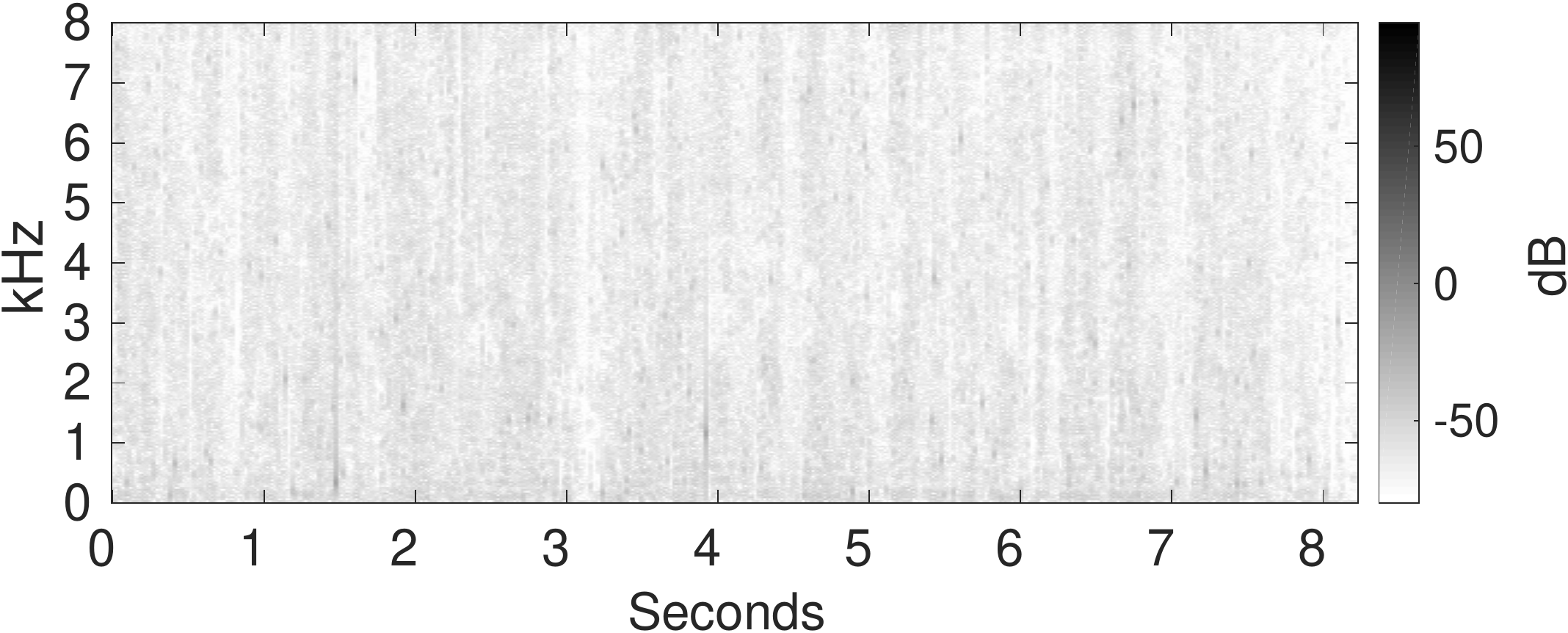}
  \caption{The power spectrum of the difference between original and adversarial~$|\mathbf D|$.}
  \label{fig:spectro_diff}
  \end{subfigure}
  \hfill
  \begin{subfigure}[t]{0.45\textwidth}
  \centering
  \includegraphics[width=\textwidth]{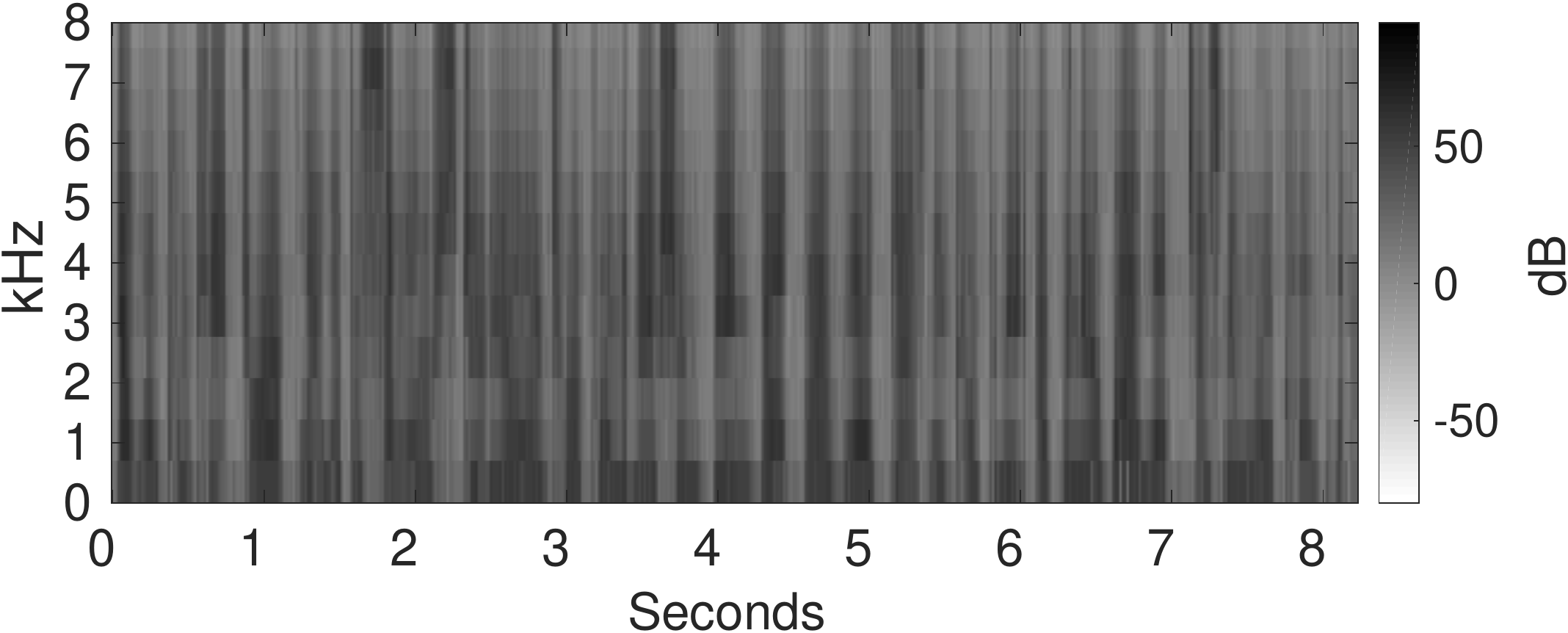}
  \caption{Hearing thresholds $\mathbf  H$.}
  \label{fig:spectro_thresholds}
  \end{subfigure}
  
  \caption{Original audio sample (\ref{fig:spectro_original}) in comparison to the adversarial audio sample (\ref{fig:spectro_spoofed}). The difference of both signals is shown in Figure~\ref{fig:spectro_diff}. Figure~\ref{fig:spectro_thresholds} visualizes the hearing thresholds of the original sample, which are used for the attack~algorithm.}
  \label{fig:spectros}
  \end{figure*}

\subsection{Hearing Thresholds}
\label{sec:hearing_thresholds}

Psychoacoustic hearing thresholds allow us to limit audible distortions from all signal manipulations. More specifically, we use the hearing thresholds during the manipulation of the input signal in order to limit audible distortions. For this purpose, we use the original audio signal to calculate the hearing thresholds~$\mathbf H$ as described in Section~\ref{sec:psychoac}. We limit the differences~$\mathbf D$ between the original signal spectrum $\mathbf S$ and the modified signal spectrum $\mathbf M$ to the threshold of human perception for all times~$t$ and frequencies~$k$
\begin{align*}
    D(t,f) & \leq H(t,k), \quad \forall t,k, \\
    \textrm{with} \quad D(t,k) &= 20 \cdot \text{log}_{10}\frac{|S(t,k)-M(t,k)|}{\text{max}_{t,k}(|\mathbf S|)}.
    \label{eq:requi}
\end{align*}
The maximum value of the power spectrum~$|\mathbf S|$ defines the reference value for each utterance, which is necessary to calculate the difference in dB. Examples for $|\mathbf  S|$, $|\mathbf  M|$, $|\mathbf  D|$, and $\mathbf  H$ in dB are plotted in Figure~\ref{fig:spectros}, where the power spectra are plotted for one utterance.

We calculate the amount of distortion that is still acceptable~via
\begin{equation}
    \mathbf \Phi = \mathbf H - \mathbf D.
 \label{eq:phi}
\end{equation}
The resulting matrix $\mathbf \Phi$ contains the difference in dB to the calculated hearing thresholds. 

In the following step, we use the matrix $\mathbf \Phi$ to derive scaling factors. First, because the thresholds are tight, an additional variable $\lambda$ is added, to allow the algorithm to differ from the hearing thresholds by small amounts
\begin{equation}
    \mathbf \Phi^* =  \mathbf \Phi + \lambda.
     \label{eq:lambda}
\end{equation}
In general, a negative value for $\Phi^* (t,k)$ indicates that we crossed the threshold.
As we want to avoid more noise for these time-frequency-bins, we set all $\Phi^* (t,k) < 0$ to zero. We then obtain a time-frequency matrix of scale factors~$ \hat{\mathbf \Phi}$ by normalizing $ \mathbf \Phi^*$ to values between zero and one, via 
\begin{equation*}
    \hat{\Phi}(t,k) =  \frac{\Phi^*(t,k) - \text{min}_{t,k}(\mathbf \Phi^*)}{\text{max}_{t,k}(\mathbf \Phi^*) - \text{min}_{t,k}(\mathbf \Phi^*)}, \quad \forall t,k.
\end{equation*}
The scaling factors are applied during each backpropagation iteration. 
Using the resulting scaling factors~$ \hat{\Phi}(t,k)$ typically leads to good results, but especially in the cases where only very small changes are acceptable, this scaling factor alone is not enough to satisfy the hearing thresholds. Therefore, we use another, fixed scaling factor, which only depends on the hearing thresholds~$\mathbf H$. For this purpose, $\mathbf H$ is also scaled to values between zero and one, denoted by $\hat{\mathbf H}$.

Therefore, the gradient~$\nabla X(t,k)$ calculated via Equation~\eqref{eq:grad_dft} between the DFT and the magnitude step is scaled by both scaling factors 
\begin{equation*}
\nabla X^*(t,k) =  \nabla X(t,k) \cdot \hat{\Phi}(t,k) \cdot \hat{H}(t,k), \quad \forall t,k.
\end{equation*}

\section{Experiments and Results}
With the help of the following experiments, we verify and assess the proposed attack. We target the ASR system \emph{Kaldi} and use it for our speech recognition experiments. We also compare the influence of the suggested improvements to the algorithm and assess the influence of significant parameter settings on the success of the adversarial attack.

\subsection{Experimental Setup}
To verify the feasibility of targeted adversarial attacks on state-of-the-art ASR systems, we have used the default settings for the \emph{Wall Street Journal} (WSJ) training recipe of the \emph{Kaldi} toolkit~\cite{Povey_ASRU2011_2011}. Only the preprocessing step was adapted for the integration into the DNN. The WSJ data set is well suited for large vocabulary ASR: it is phone-based and contains more than $80$ hours of training data, composed of read sentences of the Wall Street Journal recorded under mostly clean conditions. Due to the large dictionary with more than $100,000$ words, this setup is suitable to show the feasibility of targeted adversarial attacks for arbitrary transcriptions.

For the evaluation, we embedded the hidden voice commands (\ie target transcription) in two types of audio data: speech and music. We collect and compare results with and without the application of hearing thresholds, and with and without the use of forced alignment. All computations were performed on a 6-core Intel Core i7-4960X processor.

\subsection{Metrics}
In the following, we describe the metrics that we used to measure recognition accuracy and to assess to which degree the perturbations of the adversarial attacks needed to exceed hearing thresholds in each of our algorithm's variants.

\subsubsection{Word Error Rate}
\label{sec:wer}
As the adversarial examples are primarily designed to fool an ASR system, a natural metric for our success is the accuracy with which the target transcription was actually recognized. For this purpose, we use the Levenshtein distance~\cite{navarro2001guided} to calculate the word error rate (WER). A dynamic-programming algorithm is employed to count the number of deleted $D$, inserted $I$, and substituted $S$ words in comparison to the total number of words $N$ in the sentence, which together allows for determining the word error rate via
\begin{equation*}
WER = \frac{D + I + S}{N}.
\label{eq:wer}
\end{equation*}
When the adversarial example is based on audio samples with speech, it is possible that the original text is transcribed instead of---or in addition to---the target transcription. Therefore, it can happen that many words are inserted, possibly even more words than contained in the target text. This can lead to WERs larger than 100\,\%, which can also be observed in Table~\ref{tab:thresh}, and which is not uncommon when testing ASR systems under highly unfavorable conditions.

\subsubsection{Difference Measure}
To determine the amount of perceptible noise, measures like the signal-to-noise-ratio (SNR) are not sufficient given that they do not represent the subjective, perceptible noise. Hence, we have used~$\mathbf \Phi$ of Equation~\eqref{eq:phi} to obtain a comparable measure of audible noise. For this purpose, we only consider values $> 0$, as only these are in excess of the hearing thresholds. This may happen when $\lambda$ is set to values larger than zero, or where changes in one frequency bin also affect adjacent bins.

We sum all values $~ \Phi(t,k) > 0$ for $t = 0, \dots, T-1$ and $k = 0, \dots, N-1$ and divide the sum by $T \cdot N$ for normalization. This value is denoted by~$\phi$. It constitutes our measure of the degree of perceptibility of noise.

\begin{figure}
  \centering
  
  \begin{subfigure}[b]{0.45\textwidth}
  \centering
  \includegraphics[width=\textwidth]{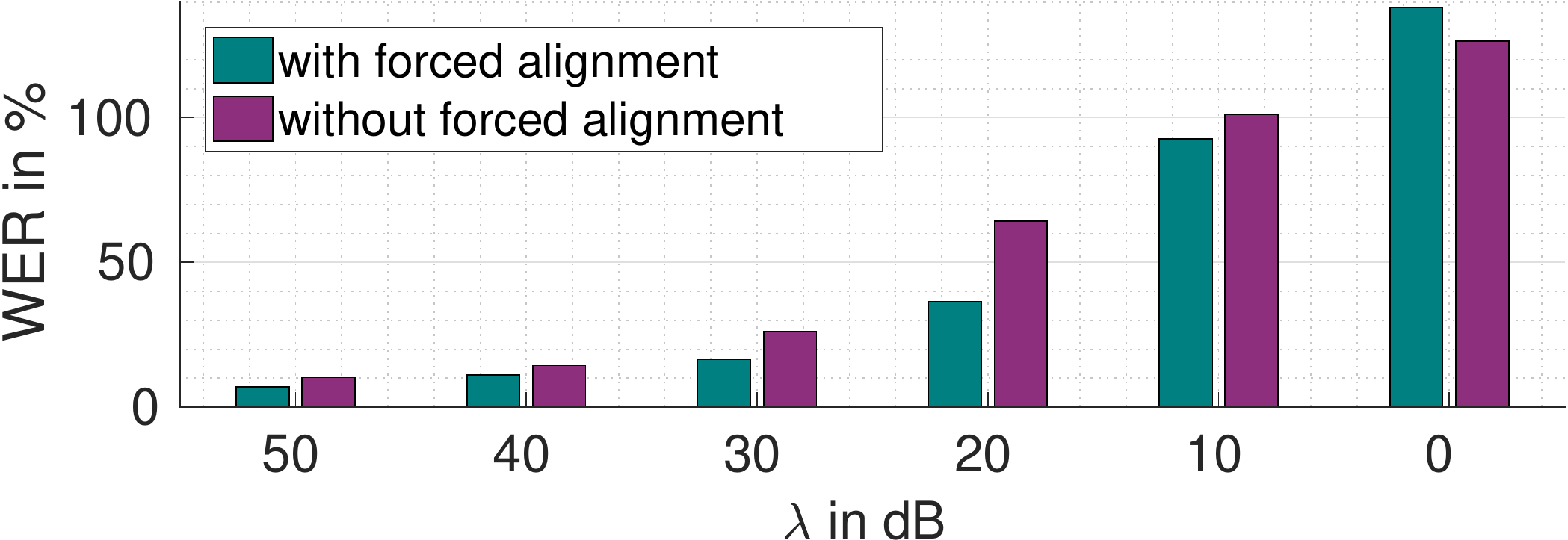}
  \label{fig:speech_0}
  \end{subfigure}%
  
  \bigskip 
  
  \begin{subfigure}[b]{0.45\textwidth}
  \centering
  \includegraphics[width=\textwidth]{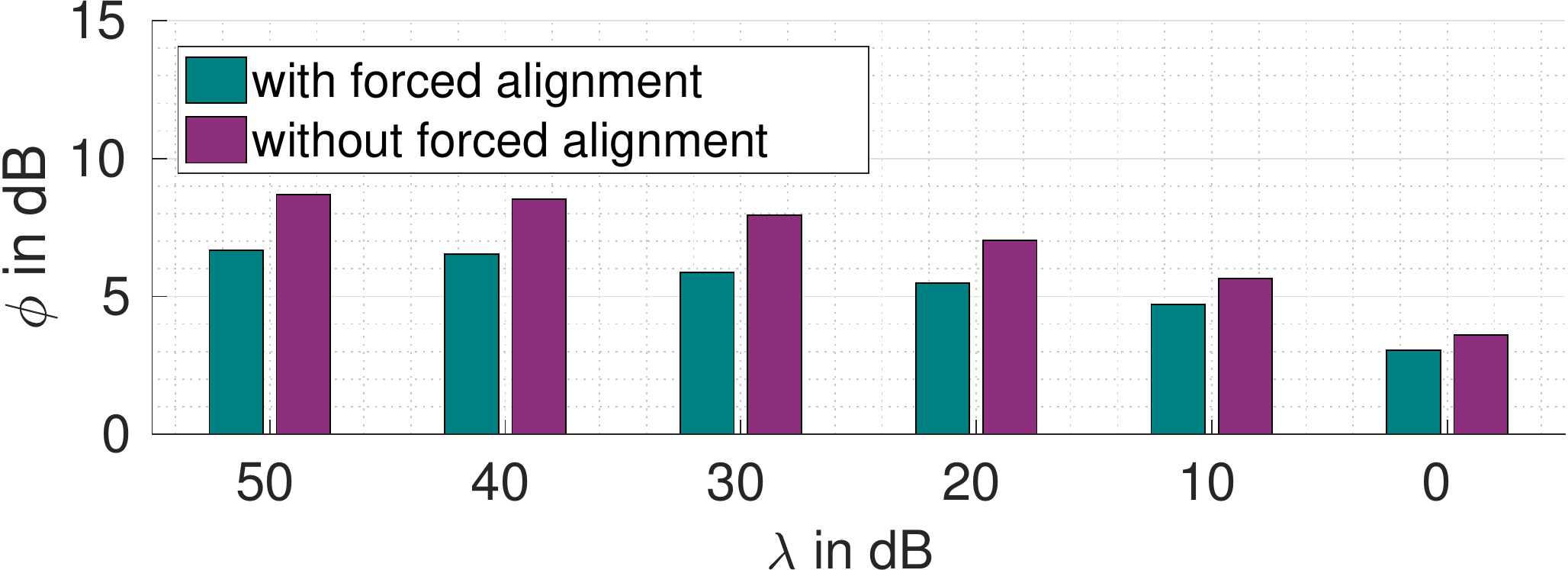}
  \label{fig:speech_20}
  \end{subfigure}%
 
\caption{Comparison of the algorithm with and without forced alignment, evaluated for different values of $\lambda$.}
  \label{fig:forced_plot}
  \end{figure}

\subsection{Improving the Attack}
As a baseline, we used a simplified version of the algorithm, forgoing both the hearing thresholds and the forced alignment stage.
In the second scenario, we included the proposed hearing thresholds. This minimizes the amount of added noise but also decreases the chance of a valid adversarial example. In the final scenario, we added the forced alignment step, which results in the full version of the suggested algorithm, with a clearly improved WER.

For the experiments, a subset of $70$ utterances for $10$ different speakers from one of the WSJ test sets was used. 

\subsubsection{Backpropagation}
First, the adversarial attack algorithm was applied without the hearing thresholds or the forced alignment. Hence, for the alignment, the audio sample was divided equally into the states of the target transcription. We used $500$ iterations of backpropagation. This gives robust results and requires a reasonable time for computation. We chose a learning rate of $0.05$, as it gave the best results during preliminary experiments. This learning rate was also used for all following experiments.
 
For the baseline test, we achieved a WER of $1.43\,\%$, but with perceptible noise. This can be seen in the average $\phi$, which was $11.62$\,dB for this scenario. This value indicates that the difference is clearly perceptible. However, the small WER shows that targeted attacks on ASR systems are possible and that our approach of backpropagation into the time domain can very reliably produce valid adversarial audio samples.
 
\subsubsection{Hearing Thresholds} \label{sssec:ht}
Since the main goal of the algorithm is the reduction of the perceptible noise, we included the hearing thresholds as described in Section~\ref{sec:hearing_thresholds}. For this setting, we ran the same test as before.
 
In this case, the WER increases to $64.29\,\%$, but it is still possible to create valid adversarial samples. On the positive side, the perceptible noise is clearly reduced. This is also indicated by the much smaller value of $\phi$ of only $7.04$\,dB.

We chose $\lambda = 20$ in this scenario, which has been shown to be a good trade-off. The choice of $\lambda$ highly influences the WER, a more detailed analysis can be found in Table~\ref{tab:thresh}.

\subsubsection{Forced Alignment}
To evaluate the complete system, we replaced the equal alignment by forced alignment. Again, the same test set and the same settings as in the previous scenarios were used. 
Figure~\ref{fig:forced_plot} shows a comparison of the algorithm's performance with and without forced alignment for different values of $\lambda$. The parameter $\lambda$ is defined in Equation~\eqref{eq:lambda} and describes the amount the result can differ from the thresholds in $dB$. As the thresholds are tight, this parameter can influence the success rate but does not necessarily increase the amount of noise. In all relevant cases, the WER and $\phi$ show better results with forced alignment. The only exception is the one case of $\lambda = 0$, where the WER is very high in all scenarios.

In the specific case of $\lambda = 20$, set as in Section \ref{sssec:ht}, a WER of $36.43$\,\% was achieved. This result shows the significant advantage of the forced alignment step. At the same time, the noise was again noticeably reduced, with $\phi = 5.49$\,dB. This demonstrates that the best temporal alignment noticeably increases the success rate in the sense of the WER, while at the same time reducing the amount of noise---a rare win-win situation in the highly optimized domain of ASR.
In Figure~\ref{fig:spectros}, an example of an original spectrum of an audio sample is compared with the corresponding adversarial audio sample. One can see the negligible differences between both signals. The added noise is plotted in Figure~\ref{fig:spectro_diff}. Figure~\ref{fig:spectro_thresholds} depicts the hearing thresholds of the same utterance, which were used in the attack algorithm.

\subsection{Evaluation}
In the next steps, the optimal settings are evaluated, considering the success rate, the amount of noise, and the time required to generate valid adversarial examples.

\subsubsection{Evaluation of Hearing Thresholds}

In Table~\ref{tab:thresh}, the results for speech and music samples are shown for $500$ and for $1000$ iterations of backpropagation, respectively. The value in the first row shows the setting of $ \lambda$. For comparison, the case without the use of hearing thresholds is shown in the column `None.' We applied all combinations of settings on a test set of speech containing $72$ samples and a test set of music containing $70$ samples. The test set of speech was the same as for the previous evaluations and the target text was the same for all audio samples.

The results in Table~\ref{tab:thresh} show the dependence on the number of iterations and on $ \lambda$. The higher the number of iterations and the higher $ \lambda$, the lower the WER becomes. The experiments with music show some exceptions to this rule, as a higher number of iterations slightly increases the WER in some cases. However, this is only true where no thresholds were employed or for $\lambda = 50$.

As is to be expected, the best WER results were achieved when the hearing thresholds were not applied. However, the results with applied thresholds show that it is indeed feasible to find a valid adversarial example very reliably even when minimizing human perceptibility.Even for the last column, where the WER increases to more than $100$\,\%, it was still possible to create valid adversarial examples, as we will show in the following evaluations.

In Table~\ref{tab:diff}, the corresponding values for the mean perceptibility~$\phi$ are shown. In contrast to the WER, the value $\phi$ decreases with $\lambda$, which shows the general success of the thresholds, as smaller values indicate a smaller perceptibility. Especially when no thresholds are used, $\phi$ is significantly higher than in all other cases. The evaluation of music samples shows smaller values of $\phi$ in all cases, which indicates that it is much easier to conceal adversarial examples in music. This was also confirmed by the listening tests~(\cf Section~\ref{sec:user_study}).

\subsubsection{Phone Rate Evaluation}

\renewcommand{\arraystretch}{1.0}
\begin{table}[t]
  \caption{WER in \% for different values for $ \lambda$ in the range of \SIrange{0}{50}{\dB}, comparing speech and music as audio inputs.}
  \centering
  \resizebox{\columnwidth}{!}{
  \begin{tabular}{lc|ccccccc}
  \toprule
  & \textbf{Iter.} & \textbf{None} & \textbf{50\,dB} & \textbf{40\,dB} & \textbf{30\,dB} & \textbf{20\,dB} & \textbf{10\,dB} & \textbf{0\,dB} \\ 
  \midrule 
  \multirow{ 2}{*}{\textbf{Speech}} & 500 & 2.14 & 6.96 & 11.07 & 16.43 & 36.43 & 92.69 & 138.21 \\ 
  & 1000 & 1.79 & 3.93 & ~5.00 & ~7.50 & 22.32 & 76.96 & 128.93 \\
  \multirow{ 2}{*}{\textbf{Music}} & 500 & 1.04 & ~8.16 & 13.89 & 22.74 & 31.77 & 60.07 & ~77.08 \\ 
  & 1000 & 1.22 & 10.07 & ~9.55 & 15.10 & 31.60 & 56.42 & ~77.60 \\ 
  \bottomrule 
  \end{tabular} 
  }
  \label{tab:thresh}
  \end{table}

  \begin{table}[h]
  \caption{The perceptibility~$\phi$ over all samples in the test sets in dB.}
  \centering
  \resizebox{\columnwidth}{!}{
  \begin{tabular}{lc|ccccccc}
\toprule
& \textbf{Iter.} & \textbf{None} & \textbf{50\,dB} & \textbf{40\,dB} & \textbf{30\,dB} & \textbf{20\,dB} & \textbf{10\,dB} & \textbf{0\,dB} \\ 
\midrule 
\multirow{ 2}{*}{\textbf{Speech}} & 500 & 10.11 & 6.67 & 6.53 & 5.88 & 5.49 & 4.70 & 3.05  \\ 
& 1000 & 10.80 & 7.42 & 7.54 & 6.85 & 6.46 & 5.72 & 3.61 \\
\multirow{ 2}{*}{\textbf{Music}} & 500 & ~4.92 & 3.92 & 3.56 & 3.53 & 3.39 & 2.98 & 2.02 \\ 
& 1000 &  ~5.03 & 3.91 & 3.68 & 3.40 & 3.49 & 3.20 & 2.30 \\ 
\bottomrule 
\end{tabular} 
}
  \label{tab:diff}
\end{table}

For the attack, timing changes are not relevant as long as the target text is recognized correctly. Therefore, we have tested different combinations of audio input and target text, measuring the number of phones that we could hide per second of audio, to find an optimum phone rate for our ASR system. For this purpose, different target utterances were used to create adversarial examples from audio samples of different lengths. The results are plotted in Figure~\ref{fig:len_plot}. For the evaluations, $500$ iterations and $\lambda = 20$ were used. Each point of the graph was computed based on $200$ adversarial examples with changing targets and different audio samples, all of them speech.

Figure~\ref{fig:len_plot} shows that the WER increases clearly with an increasing phone rate. We observe a minimum for $4$ phones per second, which does not change significantly at a smaller rate. As the time to calculate an adversarial sample increases with the length of the audio sample, $4$ phones per second is a reasonable choice.

\subsubsection{Number of Required Repetitions}
\label{sec:repeti}

\begin{figure}[b]
  \centering
  \includegraphics[width=0.45\textwidth]{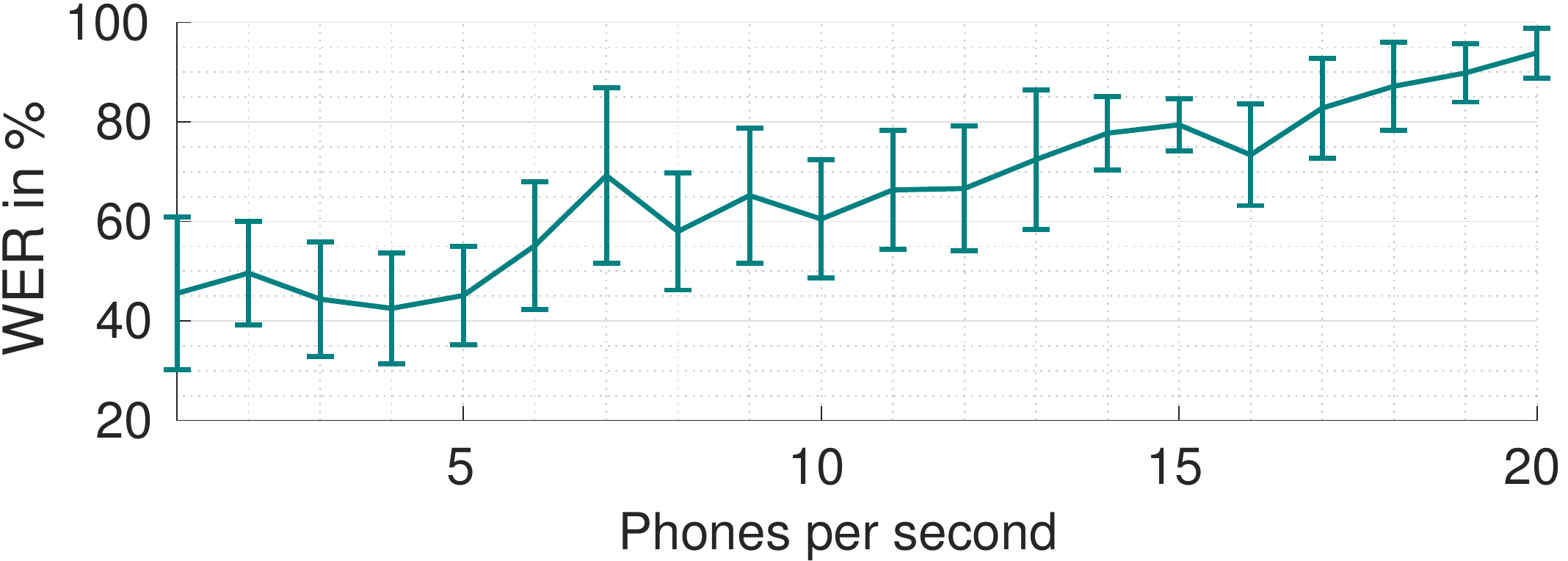}
  \caption{Accuracy for different phone rates. To create the examples, $500$ iterations of backpropagation and $\lambda = 20$ are used. The vertical lines represent the variances.}
  \label{fig:len_plot}
\end{figure}

We also analyzed the number of iterations needed to obtain a successful adversarial example for a randomly chosen audio input and target text. The results are shown in Figure~\ref{fig:success}. We tested our approach for speech and music, setting $\lambda = 0$, $\lambda = 20$, and $\lambda = 40$, respectively. For the experiments, we randomly chose speech files from $150$ samples and music files from $72$ samples. For each sample, a target text was chosen randomly from $120$ predefined texts. The only constraint was that we used only audio-text-pairs with a phone rate of \SI{6} phones per second or less, based on the previous phone rate evaluation. In the case of a higher phone rate, we chose a new audio file. We repeated the experiment $100$ times for speech and for music and used these sets for each value of $\lambda$.

For each round, we ran $100$ iterations and checked the transcription. If the target transcription was not recognized successfully, we started the next $100$ iterations and re-checked, repeating until either the maximum number of $5000$ iterations was reached or the target transcription was successfully recognized.  An adversarial example was only counted as a success if it had a WER of $0$\,\%. There were also cases were no success was achieved after $5000$ iterations. This varied from only $2$~cases for speech audio samples with $\lambda=40$ up to $9$~cases for music audio samples with $\lambda = 0$.

In general, we can not recommend using very small values of $\lambda$ with too many iterations, as some noise is added during each iteration step and the algorithm becomes slower. Even though the results in Figure~\ref{fig:success} show that it is indeed possible to successfully create adversarial samples with $\lambda$ set to zero, but $500$ or $1000$ iterations may be required. Instead, to achieve a higher success rate, it is more promising to switch to a higher value of $\lambda$, which often leads to fewer distortions overall than using $\lambda = 0$ for more iterations. This will also be confirmed by the results of the user study, which are presented in Section~\ref{sec:user_study}. 

The algorithm is easy to parallelize and for a ten-second audio file, it takes less than two minutes to calculate the adversarial perturbations with $500$ backpropagation steps on a 6-core (12 threads) Intel Core i7-4960X processor.

\begin{figure}
  \centering
  
  \begin{subfigure}[t]{0.45\textwidth}
  \centering
  \includegraphics[width=\textwidth]{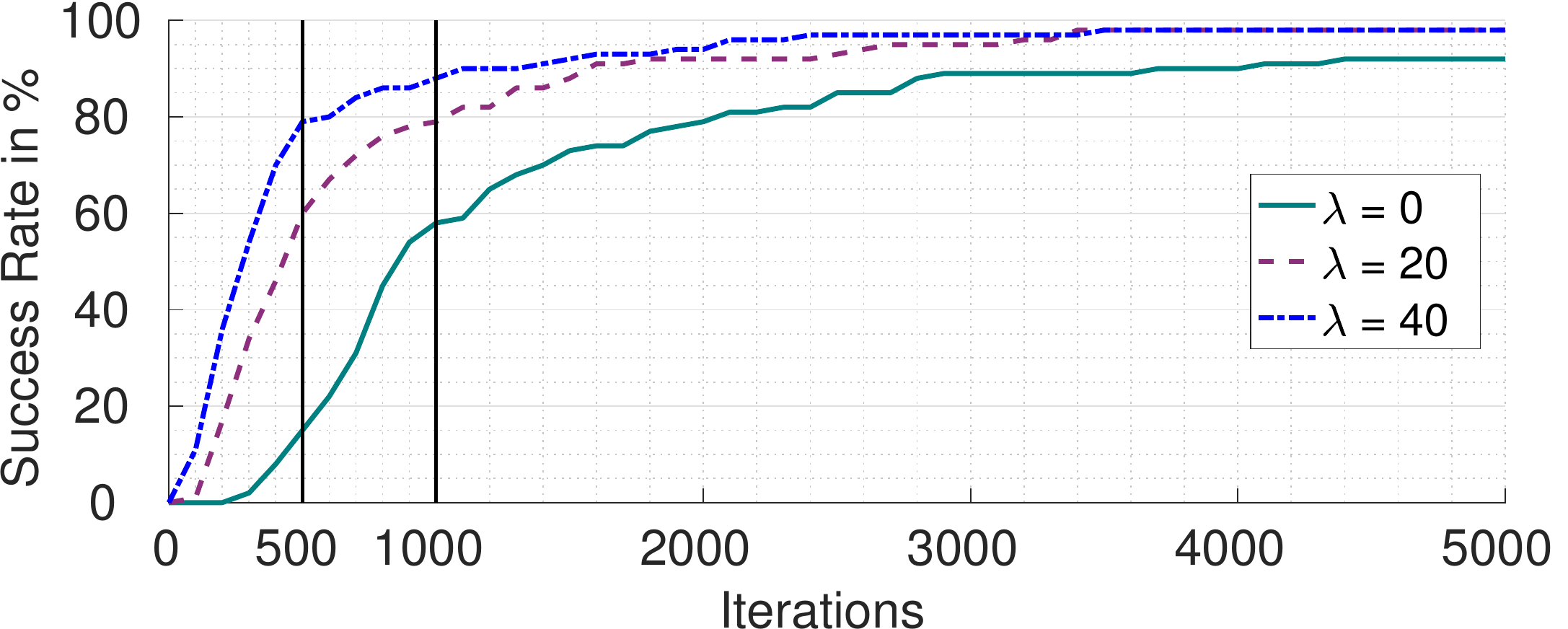}
  \caption{Speech}
  \label{fig:speech}
  \end{subfigure}%
  
  \bigskip 
  
  \begin{subfigure}[t]{0.45\textwidth}
  \centering
  \includegraphics[width=\textwidth]{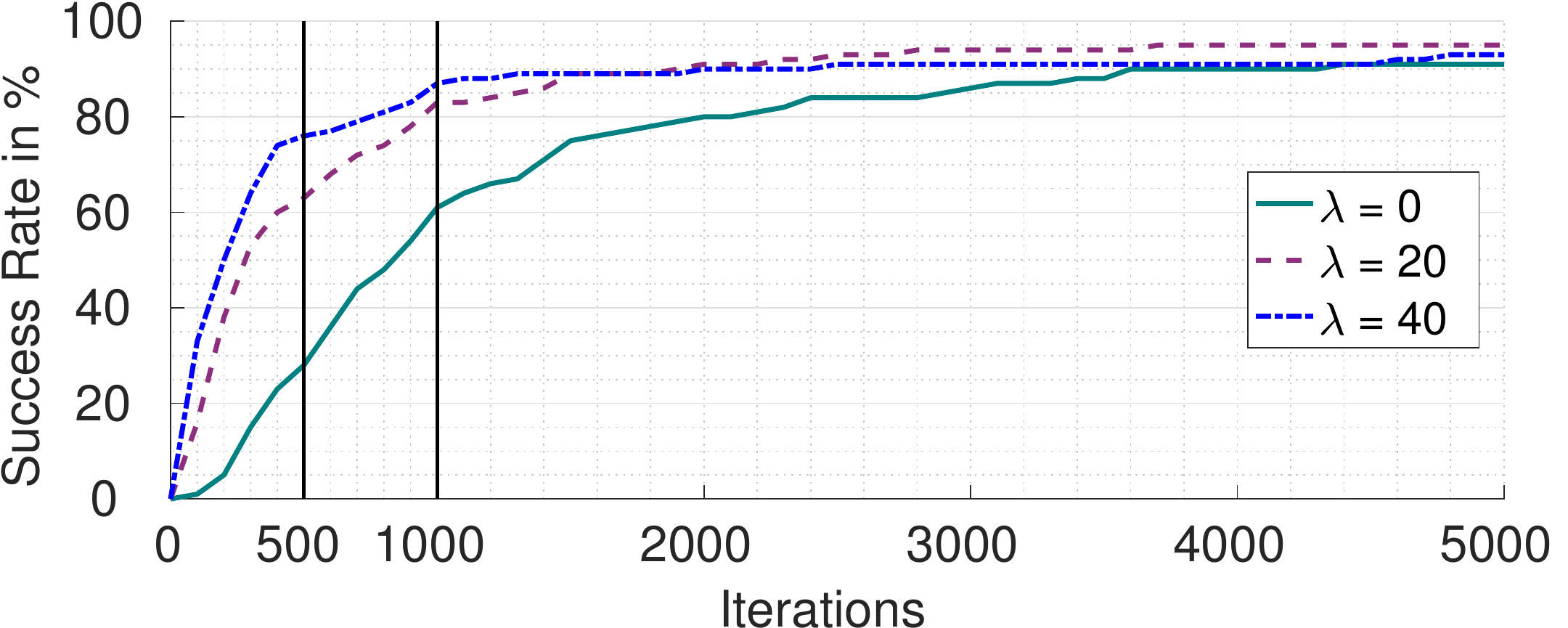}
  \caption{Music}
  \label{fig:music}
  \end{subfigure}%
 
  \caption{Success rate as a function of the number of iterations. The upper plot shows the result for speech audio samples and the bottom plot the results for music audio samples. Both sets were tested for different settings of $\lambda$.}
  \label{fig:success}
  \end{figure}

\subsection{Comparison}
 
We compare the amount of noise with \emph{CommanderSong}~\cite{yuan-18-commandersong}, as their approach is also able to create targeted attacks using \emph{Kaldi} and therefore the same DNN-HMM-based ASR system. Additionally, is the only recent approach, which reported she signal-to-noise-ratio~(SNR) of their results.

The SNR measures the amount of noise~$\sigma$, added to the original signal~$x$, computed via

\begin{equation*}
\text{SNR(dB)} = 10 \cdot \text{log}_{10}\frac{P_x}{P_{\sigma}} ,
\end{equation*}

where $P_x$ and $P_{\sigma}$ are the energies of the original signal and the noise. This means, the \emph{higher} the SNR, the \emph{less} noise was added.

Table~\ref{tab:compare} shows the SNR for successful adversarial samples, where no hearing thresholds are used (None) and for different values of $\lambda$ (40\,dB, 20\,dB, and 0\,dB) in comparison to \emph{CommanderSong}. Note, that the SNR does not measure the perceptible noise and therefore, the resulting values are not always consistent with the previously reported $\phi$. Nevertheless, the results show, that in all cases, even if no hearing thresholds are used, we achieve higher SNRs, meaning, less noise was added to create a successful adversarial example. 

\begin{table}[t]
  \caption{Comparison of SNR with \emph{CommanderSong}~\cite{yuan-18-commandersong}, best result shown in bold print.}
  \centering
  \resizebox{\columnwidth}{!}{
  \begin{tabular}{c|cccc|c}
\toprule
& \textbf{None} & \textbf{40\,dB} & \textbf{20\,dB} & \textbf{\textbf{0\,dB}} & \textbf{\emph{CommanderSong}~\cite{yuan-18-commandersong}} \\ 
\midrule 
\textbf{SNR} & 15.88 & 17.93 & \textbf{21.76} & 19.38 & 15.32 \\ 
\bottomrule 
\end{tabular} 
}
  \label{tab:compare}
\end{table}

\section{User Study}\label{sec:user_study}
We have evaluated the human perception of our audio manipulations through a two-part user study. In the transcription test, we verified that it is impossible to understand the voice command hidden in an audio sample. The MUSHRA test provides an estimate of the perceived audio quality of adversarial examples, where we tested different parameter setups of the hiding process.

\subsection{Transcription Test}
While the original text of a speech audio sample should still be understandable by human listeners, we aim for a result where the hidden command cannot be transcribed or even identified as speech. Therefore, we performed the \emph{transcription test}, in which test listeners were asked to transcribe the utterances of original and adversarial audio samples.

\subsubsection{Study Setup}
Each test listener was asked to transcribe $21$ audio samples. The utterances were the same for everyone, but with randomly chosen conditions:  $9$ original utterances, $3$ adversarial examples with $\lambda = 0$, $\lambda = 20$, and $\lambda = 40$ respectively and $3$ difference signals of the original and the adversarial example, one for each value of $\lambda$.  For the adversarial utterances, we made sure that all samples were valid, such that the target text was successfully hidden within the original utterance. We only included adversarial examples which required $\leq 500$ iterations.

We conducted the tests in a soundproofed chamber and asked the participants to listen to the samples via headphones. The task was to type all words of the audio sample into a blank text field without any provision of auto-completion, grammar, or spell checking. Participants were allowed to repeat each audio sample as often as needed and enter whatever they understood. In a post-processing phase, we performed manual corrections on minor errors in the documented answers to address typos, misspelled proper nouns, and numbers. We provide an example of the post-processing in Appendix~\ref{app:postpro}.
After revising the answers in the post-processing step, we calculated the WER using the same algorithms as introduced in Section~\ref{sec:wer}.

\subsubsection{Results}

\begin{figure}
  \centering
  \includegraphics[width=0.5\textwidth]{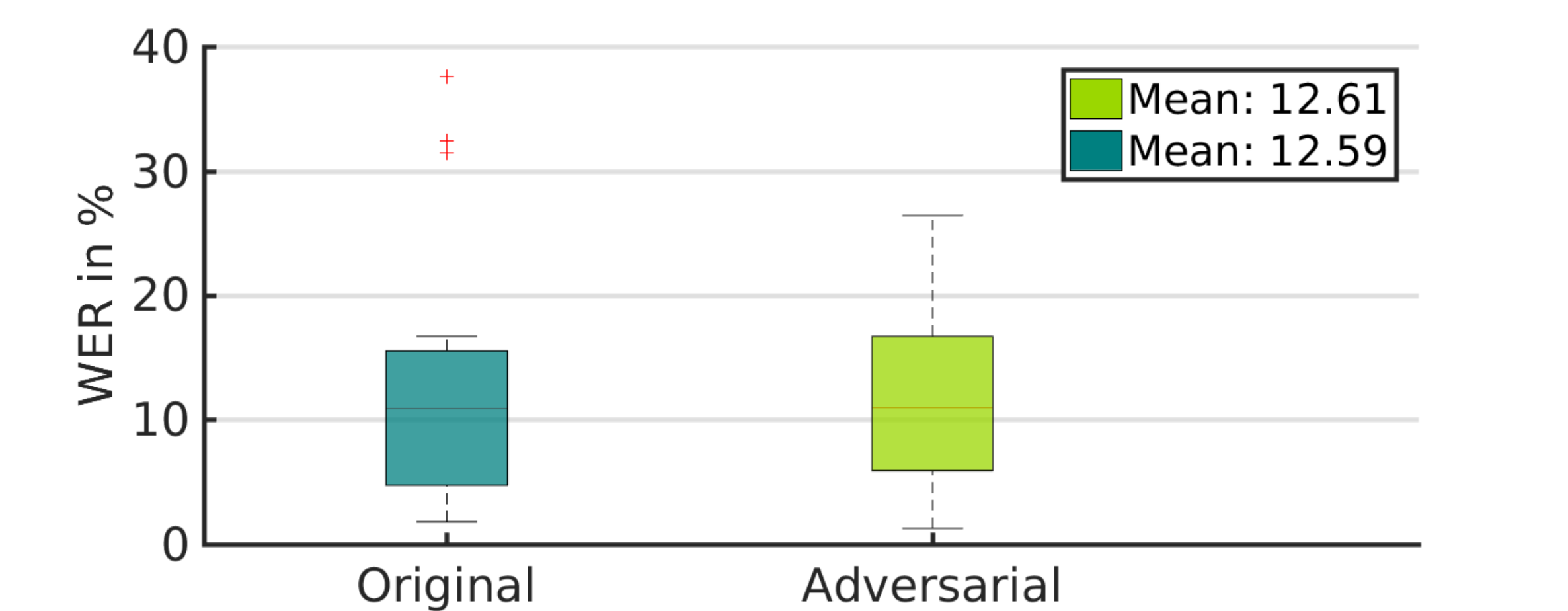}
  \caption{WER for all \num{21} utterances over all test listeners of the original utterances and the adversarial utterances.}
  \label{fig:transbox}
\end{figure}

For the evaluation, we have collected data from \num{22} listeners during an internal study at our university. None of the listeners were native speakers, but all had sufficient English skills to understand and transcribe English utterances. As we wanted to compare the WER of the original utterances with the adversarial ones, the average WER of \SI{12.52}{\percent} overall test listeners was sufficient. This number seems high, but the texts of the WSJ are quite challenging. All original transcriptions and target transcriptions are presented in Appendix~\ref{app:tranription}.
For the evaluation, we ignored all cases where only the difference of the original and adversarial sample was played. For all of these cases, none of the test listeners was able to recognize any kind of speech and therefore no text was transcribed.

For the original utterances and the adversarial utterances, an average WER of \SI{12.59}{\percent} and \SI{12.61}{\percent} was calculated. The marginal difference shows that the difference in the audio does not influence the intelligibility of the utterances.
Additionally, we have tested the distributions of the original utterances and the adversarial utterances with a two-sided t-test to verify whether both distributions have the same mean and variance. The test with a significance level of \SI{1}{\percent} shows no difference for the distributions of original and adversarial utterances. 

In the second step, we have also compared the text from the test listeners with the text which was hidden in the adversarial examples. For this, we have measured a WER far above $100$\,\%, which shows that the hidden text is not intelligible. Also, there are only correct words which were in the original text, too, and in all cases these were frequent, short words like \emph{is}, \emph{in}, or \emph{the}.

\subsection{MUSHRA Test}
In the second part of the study, we have conducted a Multiple Stimuli with Hidden Reference and Anchor (MUSHRA) test, which is commonly used to rate the quality of audio signals~\cite{schinkel2013audio}.

\subsubsection{Study Setup}
The participants were asked to rate the quality of a set of audio signals with respect to the original signal. The set contains different versions of the original audio signal under varying conditions. As the acronym shows, the set includes a \emph{hidden reference} and an \emph{anchor}. The former is the sample with the best and the latter the one with the worst quality. In our case, we have used the original audio signal as the \emph{hidden reference} and the adversarial example, which was derived without considering the hearing thresholds, as \emph{anchor}. 
Both the \emph{hidden reference} and the \emph{anchor} are used to exclude participants, who were not able to identify either the \emph{hidden reference} or the  \emph{anchor}. As a general rule, the results of participants who rate the \emph{hidden reference} with less than \num{90} MUSRHA-points more than \SI{15}{\percent} of the time are not considered. Similarly, all results of listeners who rate the \emph{anchor} with more than \num{90} MUSRHA-points more than \SI{15}{\percent} of the time are excluded.
We used the \emph{webMUSHRA} implementation, which is available online and was developed by AudioLabs~\cite{schoeffler2018webmushra}. 
 
We have prepared a MUSHRA test with nine different audio samples, three for speech, three for music, and three for recorded twittering birds. For all these cases, we have created adversarial examples for $\lambda = 0$, $\lambda = 20$,  $\lambda = 40$, and without hearing thresholds. Within one set, the target text remained the same for all conditions, and in all cases, all adversarial examples were successful with $\leq 500$ iterations. 
The participants were asked to rate all audio signals in the set on a scale between 0 and 100 (0--20: Bad, 21--40: Poor, 41--60: Fair, 61--80: Good, 81--100: Excellent).
Again, the listening test was conducted in a soundproofed chamber and via headphones.

\subsubsection{Results}
\begin{figure}
  \centering
  
  \begin{subfigure}{0.49\textwidth}
  \centering
  \includegraphics[width=\textwidth]{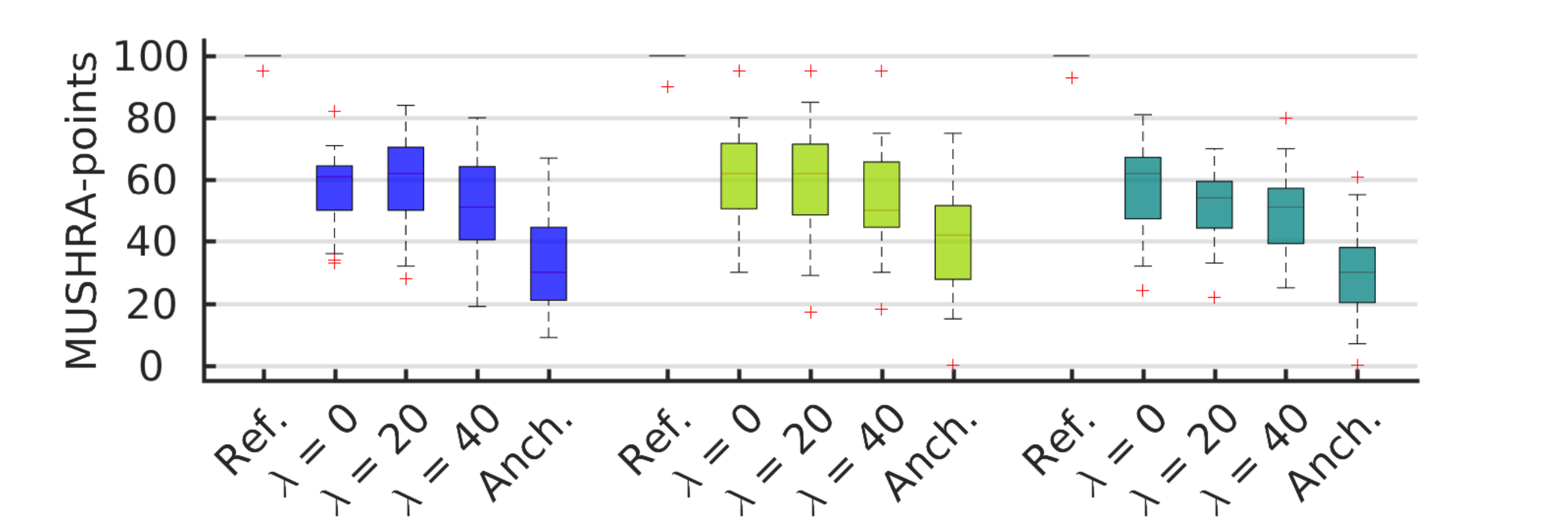}
  \caption{Speech}
  \label{fig:speech_box}
  \end{subfigure}%
  \begin{subfigure}{0.49\textwidth}
  \centering
  \includegraphics[width=\textwidth]{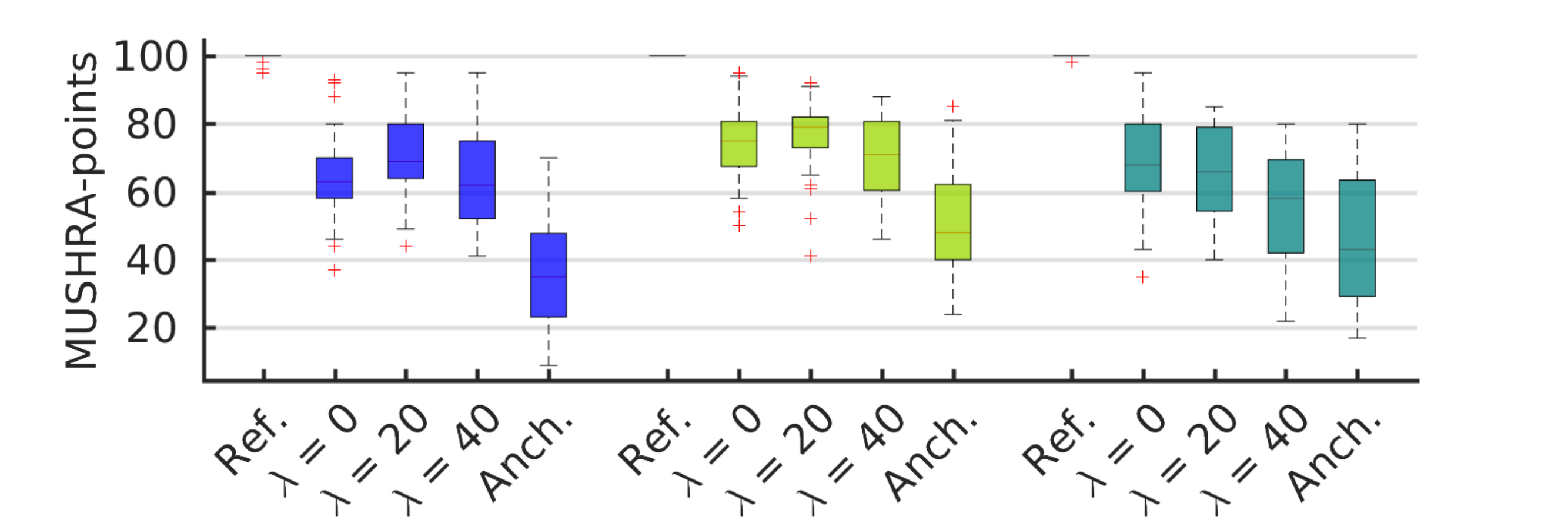}
  \caption{Music}
  \label{fig:music_box}
  \end{subfigure}

  \begin{subfigure}{0.49\textwidth}
  \centering
  \includegraphics[width=\textwidth]{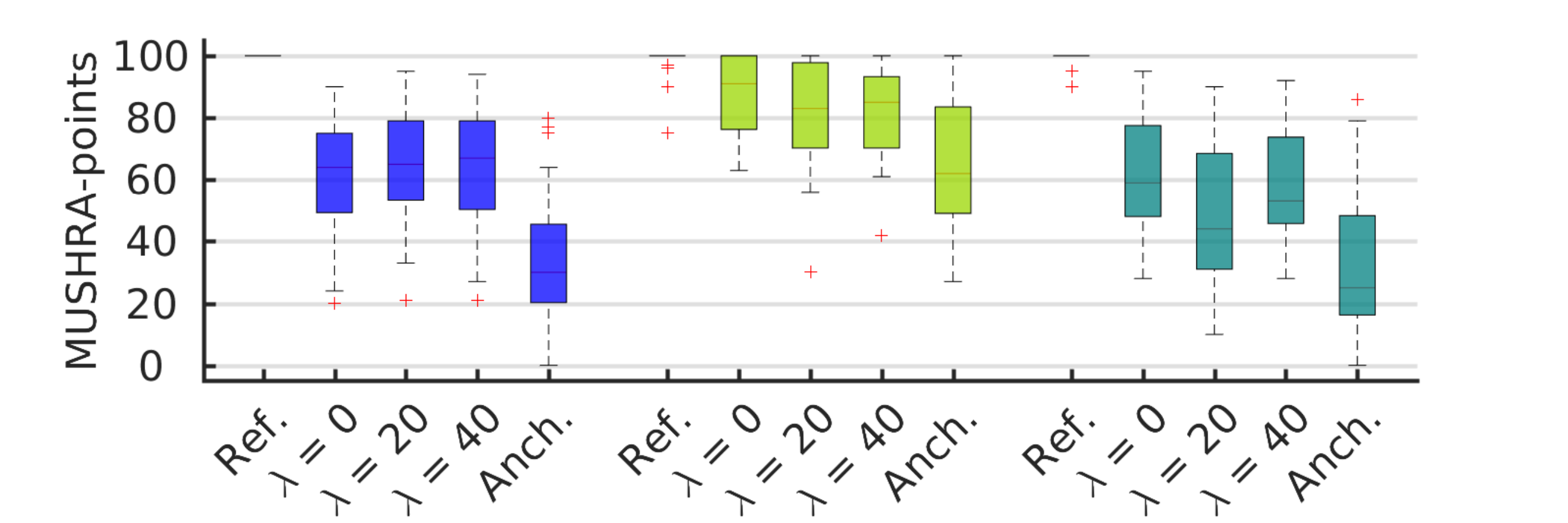}
  \caption{Birds}
  \label{fig:birds_box}
  \end{subfigure}%

  \caption{Ratings of all test listeners in the MUSHRA test. We tested three audio samples for speech, music, and twittering birds. The left box plot of all nine cases shows the rating of the original signal and therefore shows very high values. The anchor is an adversarial example of the audio signal that had been created without considering hearing thresholds.}
  \label{fig:mushra}
  \end{figure}
  
We have collected data from $30$ test listeners, $3$ of whom were discarded due to the MUSHRA exclusion criteria. The results of the remaining test listeners are shown in Figure~\ref{fig:mushra} for all nine MUSHRA tests. In almost all cases, the reference is rated with $100$ MUSHRA-points. Also, the anchors are rated with the lowest values in all cases.

We tested the distributions of the anchor and the other adversarial utterances in one-sided t-tests. For this, we used all values for one condition overall nine MUSHRA tests. The tests with a significance level of \SI{1}{\percent} show that in all cases, the anchor distribution without the use of hearing thresholds has a significantly lower average rating than the adversarial examples where the hearing thresholds are used. Hence, there is a clear perceptible difference between adversarial examples with hearing thresholds and adversarial examples without hearing thresholds.

During the test, the original signal was normally rated higher than the adversarial examples. However, it has to be considered that the test listeners directly compared the original signal with the adversarial ones. In an attack scenario, this would not be the case, as the original audio signal is normally unknown to the listeners. Despite the direct comparison, there is one MUSRHA test where the adversarial examples with hearing thresholds are very frequently rated with a similar value as the reference and more than $80$ MUSHRA-points. This is the case for the second test with twittering birds, which shows that there is a barely perceptible difference between the adversarial samples and the original audio signal.

Additionally, we observed that there is no clear preference for a specific value of $\lambda$. The samples with $\lambda = 0$ received a slightly higher average rating in comparison to  $\lambda = 20$ and $\lambda = 40$, but there is only a significant difference for the distributions of $\lambda = 0$ and $\lambda = 40$. This can be explained with the different number of iterations, since, as shown in Section~\ref{sec:repeti}, for a higher value of $\lambda$, fewer iterations are necessary and each iteration can add noise to the audio signal.

\section{Related Work}
Adversarial machine learning techniques have seen a rapid development in the past years, in which they were shown to be highly successful for image classifiers. Adversarial examples have also been used to attack ASR systems, however, there, the modifications of the signals are usually quite perceptible (and sometimes even understandable) for human listeners. In the following, we review existing literature in this area and discuss the novel contributions of our approach.

\subsection{Adversarial Machine Learning Attacks}
There are many examples of successful adversarial attacks on image files in the recent past and hence we only discuss selected papers. In most cases, the attacks were aimed at classification only, either on computer images or real-world attacks. 
For example, Evtimov \etal showed one of the first real-world adversarial attacks~\cite{evtimov2017robust}. They created and printed stickers, which can be used to obfuscate traffic signs. For humans, the stickers are visible. However, they seem very inconspicuous and could possibly fool autonomous cars.
Athalye and Sutskever presented another real-world adversarial perturbation on a 3D-printed turtle, which is recognized as a rifle from almost every point of view~\cite{athalye2017synthesizing}. The algorithm to create this 3D object not only minimizes the distortion for one image, but for all possible projections of a 3D object into a 2D image.
A similar attack on a universal adversarial perturbation was presented by Brown \etal~\cite{brown2017adversarial}. They have created a patch which works universally and can be printed with any color printer. The resulting image will be recognized as a toaster without covering the real content even partially.
An approach which works for tasks other than classification is presented by Cisse \etal~\cite{cisse2017houdini}. The authors used a probabilistic method to change the input signal and also showed results for different tasks but were not successful in implementing a robust targeted attack for an ASR system.
Carlini \etal introduced an approach with a minimum of distortions where the resulting images only differ in a few pixels from the original files~\cite{carlini2017towards}. Additionally, they are robust against common distillation defense techniques~\cite{papernot2016distillation}.

Compared to attacks against audio signals, attacks against image files are easier, as they do not have temporal dependencies. Note that the underlying techniques for our attack are similar, but we had to refine them for the audio domain.

\subsection{Adversarial Voice Commands}

Adversarial attacks on ASR systems focus either on hiding a target transcription~\cite{carlini2016hidden} or on obfuscating the original transcription~\cite{cisse2017houdini}. Almost all previous works on attacks against ASR systems were not DNN-based and therefore use other techniques~\cite{carlini2016hidden,zhang2017dolphinattack,vaidya2015cocaine}. Furthermore, none of the existing attacks used psychoacoustics to hide a target transcription in another audio signal.

Carlini \etal have shown that targeted attacks against HMM-only ASR systems are possible~\cite{carlini2016hidden}. They use an inverse feature extraction to create adversarial audio samples. However, the resulting audio samples are not intelligible by humans in most of the cases and may be considered as noise, but may make thoughtful listeners suspicious.
A different approach was shown by Vaidya \etal~\cite{vaidya2015cocaine}, where the authors changed an input signal to fit the target transcription by considering the features instead of the output of the DNN. Nevertheless, the results show high distortions of the audio signal and can easily be detected by a human.

An approach to overcome this limitation was proposed by Zhang \etal They have shown that an adversary can hide a transcription by utilizing non-linearities of microphones to modulate the baseband audio signals with ultrasound above 20\,kHz~\cite{zhang2017dolphinattack}. The main downside of the attack is the fact that the information of the necessary features needs to be retrieved from audio signal, recorded with the specific microphone, which is costly in practice. Furthermore, the modulation is tailored to a specific microphone an adversary wants to attack. As a result, the result may differ if another microphone is used. Song and Mittael~\cite{song2017inaudible} and Roy \etal~\cite{roy2017backdoor} introduced similar ultrasound-based attacks that are not adversarial examples, but rather interact with the ASR system in a frequency range inaudible to humans.

Recently and concurrently, Carlini and Wagner published a technical report in which they introduce a general targeted attack on ASR systems using CTC-loss~\cite{carlini2018audio}. The attack is based on a gradient-descent-based minimization~\cite{carlini2017towards} (as used in previous image classification adversarial attacks), but the loss function is represented via CTC-loss, which is optimized for time sequences. Compared to our approach, the perceptible noise level is higher and the attack is less effective, given that the algorithm needs a lot of time to calculate an adversarial example since it is based on a grid search.

\emph{CommanderSong}~\cite{yuan-18-commandersong} is also evaluated against Kaldi and uses backpropagation to find an adversarial example. However, in order to minimize the noise, approaches from the image domain are borrowed. Therefore, the algorithm does not consider human perception. Additionally, the attack is only shown for music and the very limited over-the-air attack highly depends on the speakers and recording devices as the attack parameters have to be adjusted especially for these components.

Our approach is different from all previous studies on adversarial perturbations for ASR, as we combine a targeted attack with the requirement that the added noise should be barely, if at all, perceptible. We use a modified backpropagation scheme, which has been very successful in creating adversarial perturbations for image classification and we initialize our optimization by forced-alignment to further minimize audible~noise.

\subsection{Hiding Information in Audio}
Watermarking approaches use human perception to hide information about an image, video, or audio clip within itself~\cite{barni-01-improved, cox-07-digital, wolfgang-99-perceptual}. The purpose in the case of watermarking, however, differs from our method and steganography, as it is used for copyright protection. Watermarking uses algorithms to hide information in audio signals within the lower frequencies or also with help of a psychoacoustic model~\cite{wu-2004-self,seok-2001-audio}. Differently from watermarking and steganography, the frequency ranges cannot be chosen arbitrarily when an ASR system is to be attacked. Instead, the information must be presented in just those frequency regions, on which the ASR has been trained.

Audio steganography is motivated by the challenge of hiding additional information in an acoustic carrier signal, \eg for transmitting sensitive information in case of comprehensive Internet censorship~\cite{skypeline}. LSB techniques~\cite{lsb3,lsb4} manipulate the binary representation of a signal and hide information in the least significant bits of a signal, which limits the perceived distortions to a minimum. In contrast to our work, such schemes ignore the acoustic characteristics of the carrier signal and achieve their hiding capabilities at the expense of disrupting the statistical characteristics of the original input. Modulation-based systems~\cite{dsss12b,skypeline} manipulate the carrier signal in the time or frequency domain to encode information within the signal characteristics. Such modulations allow the attacker to consider the frequency or energy features of a signal and help to provide a less conspicuous manipulation of the carrier signal. Both classes of steganography systems aim at hiding information in a legitimate carrier signal but are focused on creating a protected transmission channel within an untrusted system. In contrast, while our work is designed to provide a comparable level of \emph{inconspicuousness}, we require the successful transcription of a target message through an automated speech recognition system, which precludes the use of watermarking or steganography algorithms here.

\section{Discussion}
We have shown that it is possible to successfully attack state-of-the-art DNN-HMM ASR systems with targeted adversarial perturbations, which are barely or even impossible to distinguish from original audio samples.
Based on different experiments, we demonstrated that it is possible to find the best setup for the proposed algorithm for the creation of adversarial examples. 
However, these results also open questions regarding possible countermeasures and future work.

\subsection{Parameter Choice}
The choice of the parameters highly affects the amount of perceptible noise. The evaluation has shown that a higher number of iterations increases the success rate, but simultaneously the amount of noise. However, for iterations $< 500$ the success rate is already very high and therefore, $500$ should not be exceeded. Additionally, by this choice, the required calculation time is reduced as well.
If the success rate needs to be raised, the increase of $\lambda$ had a higher effect. Although the participants in the MUSHRA test did prefer smaller values for $\lambda$, there was no significant difference if $\lambda$ was only increased by $20$\,dB. 
Additionally, the phone rate should be set to an optimum value as this highly affects the success of the attack.

Besides improving the success of the attack, the choice of the original audio sample greatly influences the quality of the adversarial example. There might be use cases, where the original audio sample is fixed, but in general, the choice of the original sample is free. We recommend using music or other unsuspicious audio samples, like bird twittering, which do not contain speech, as speech has to be obfuscated, typically leading to larger required adversarial perturbations.

The process can be parallelized and is relatively fast in comparison to other attacks proposed in the past, as we have integrated the preprocessing into the backpropagation. Therefore, we recommend to use different promising setups and to choose that one which sounds the most inconspicuous while giving the required success rate.

\subsection{Countermeasures}
Distillation was shown to be a successful countermeasure against attacks in image classification~\cite{papernot2016distillation}. It is a technique to improve the robustness of classification-based DNNs~\cite{hinton2015distilling}, which uses the output of a pre-trained DNN as soft labels in order to train a second DNN with these soft labels as targets. However, the transcription of the DNN-based ASR system not only depends on the classification result but also on the temporal alignment. Therefore, distillation might not be an appropriate countermeasure for ASR systems.

A general countermeasure could be to consider human perception. A very simple version would be to apply MP3 encoding to the input data. However, the DNN is not trained for that kind of data. Nevertheless, we did run some tests on our adversarial examples. With this setup, the original transcription could not be recovered, but the target transcription was also distorted.
We assume that training the ASR-DNN with MP3-encoded audio files will only move the vulnerability into the perceptible region of the audio files, but will not circumvent blind spots of DNNs completely.

\subsection{Future Work}
One obvious question is whether our attack also works in a real-world setup. For real-world attacks, different circumstances have to be considered. This mainly includes the acoustic transfer function, which is relevant if an audio signal is transferred over the air. Also, additional noise from different sources can be present in a real-world environment. 
In a controlled environment (\eg an elevator) it is possible to calculate the transfer function and to consider or exclude external noise. Additionally, it is not unusual to have music in an elevator, which would make such a setup very unsuspicious. 

A similar attack can also be imagined for real applications, \eg Amazon's Alexa. However, the detailed architecture of this system is hard to access and it requires a-priori investigations to obtain that kind of information. It may be possible to retrieve the model parameters with different model stealing approaches~\cite{tramer2016stealing,papernot2017blackbox,wang2018hyperparameter,papernot2016transferability,juuti2018prada}. For Alexa, our reverse engineering results of the firmware indicate that Amazon uses parts of \emph{Kaldi}. Therefore, a limited knowledge about the topology and parameters might be enough to create a model for a black-box attack. A starting point could be the keyword recognition of commercial ASR systems, \eg ``\emph{Alexa}.'' This would have the advantage that the keyword recognition runs locally on the device and would, therefore, be easier to access. 

For image classification, universal adversarial perturbations have already been successfully created~\cite{brown2017adversarial,moosavi2017universal}. For ASR systems, it is still an open question if these kinds of adversarial perturbations exist and how they can be created. 

\section{Conclusion}
We have presented a new method for creating adversarial attacks on ASR systems, which explicitly take dynamic human hearing thresholds into account. In this way, borrowing the mechanisms of MP3 encoding, the audibility of the added noise is clearly reduced. 
We perform our attack against the state-of-the-art \emph{Kaldi} ASR system and feed the adversarial input directly into the recognizer in order to show the general feasibility of psychoacoustics-based attacks.

By applying forced alignment and backpropagation to the DNN-HMM system, we were able to create inconspicuous adversarial perturbations very reliably. In general, it is possible to hide any target transcription within any audio file and, with the correct attack vectors, it was possible to hide the noise below the hearing threshold and make the changes psychophysically almost imperceptible.
The choice of the original audio sample, an optimal phone rate, and forced alignment give the optimal starting point for the creation of adversarial examples. Additionally, we have evaluated different algorithm setups, including the number of iterations and the allowed deviation from the hearing thresholds. The comparison with another approach in~\cite{yuan-18-commandersong}, which is also able to create targeted adversarial examples, shows that our approach needs far lower distortions.
Listening tests have proven that the target transcription was incomprehensible for human listeners. Furthermore, for some audio files, it was almost impossible for participants to distinguish between the original and adversarial sample, even with headphones and in a direct comparison.

Future work should investigate the hardening of ASR systems by considering psychoacoustic models, in order to prevent these presently fairly easy attacks. Additionally, similar attacks should be evaluated on commercial ASR systems and in real-world attacks with a black-box setting.

%
%

\bibliography{references}

\onecolumn
\appendix
\subsection{Example of Post-Processing of One Entire Transcription Listening Test}
\label{app:postpro}
\scriptsize
\begin{verbatim}
01 original: QUOTE AN EYE FOR AN EYE END QUOTE
01 modified: "QUOTE AN EYE FOR AN EYE END QUOTE

02 original: KAISERTEK DIDN'T NAME SUCCESSORS FOR MISTER MAYER AT ITS POST
02 modified: KAISERTECH DIDN'T NAME SUCCESSORS FOR MR. MAIER AT ITS POST

03 original: BUT ROBERT UNDERWOOD PRESIDENT OF THE GOLDEN GOLF SUBSIDIARY SAID THERE ARE PERFECTLY NATURAL SYNERGIES...
03 modified: BUT ROBERT UNDERWOOD PRESIDENT OF THE GOLDEN GULF SUBSIDIARY SAID THERE ARE PERFECTLY NATURAL SYNERGIES... 
03 original: ...BETWEEN THE BUSINESSES
03 modified: ...BETWEEN THE BUSINESSES

04 original: THEY NOTED THAT TIME SQUARE AREA HAS BEEN SLOWLY IMPROVING BECAUSE THE MARKET FOR
04 modified: THEY NOTED THAT TIME SQUARE AREA HAS BEEN SLOWLY IMPROVING BECAUSE THE MARKET FOR

05 original: BUT I HAVE TO LIVE WITH MYSELF
05 modified: BUT I HAVE TO LIVE WITH MYSELF

06 original: XXX SAID ITS CHAIRMAN AND CHIEF EXECUTIVE OFFICER PETER J POPE MADE THE EARNINGS PROJECTION... 
06 modified: POPE  AND  TALBOT SAID ITS CHAIRMAN AND CHIEF EXECUTIVE OFFICER PETER T. POPE MADE THE EARNINGS PROJECTION... 
06 original: ...TO THE NEW YORKS SECURITY ANALYSTS
06 modified: ...TO THE NEW YORKS SECURITY ANALYSTS

07 original: CONSUMER SPENDING SEARCH POINT SINCE JUNE LET BY AN JUMP IN AUTO SALES
07 modified: CONSUMER SPENDING SEARCH POINT SINCE JUNE LED BY AN JUMP IN AUTO SALES

08 original: BUT NOW THE SOCIAL CONTRACT IS CRACKING
08 modified: BUT NOW THE SOCIAL CONTRACT IS CRACKING

09 original: WILLIAMS LYMAN AND BLOOMENBAUM MR MILKAN HAS RETAINED AN EXPERIENCED TEAM
09 modified: WILLIAMS LIMAN AND FLUMENBAUM MR. MILKEN HAS RETAINED AN EXPERIENCED TEAM

10 original: ----------------------------------------------
10 modified: 

11 original: THE NINETEENTH SUMMARY ANUAL REPORT PRODUCED BY THE PARTICIPANTS IN THE FINANCIAL EXECUTIVE INSTITUTE... 
11 modified: THE NINETEENTH SUMMARY ANNUAL REPORT PRODUCED BY THE PARTICIPANTS IN THE FINANCIAL EXECUTIVE INSTITUTE... 
11 original: ...PROJECT SHARPLY REDUCED THE NUMBER OF PAGES OF FINANCIAL STATEMENTS TYPICALLY  FROM TWENTYTWO TO TO TEN
11 modified: ...PROJECT SHARPLY REDUCED THE NUMBER OF PAGES OF FINANCIAL STATEMENTS TYPICALLY  FROM TWENTY TWO TO TEN

12 original: A THIRD MADE SIGNIFICANT CONTENTS CHANGES SUCH AS MOVING FOOTNOTE MATERIAL INTO THE NARRATIVE USING GRAPHS... 
12 modified: A THIRD MADE SIGNIFICANT CONTENTS CHANGES SUCH AS MOVING FOOTNOTE MATERIAL INTO THE NARRATIVE USING GRAPHS... 
12 original: ...MORE GENEROUSLY AND REWRITING THE FINANCIAL REVIEW IN LEYMANS TERMS
12 modified: ...MORE GENEROUSLY AND REWRITING THE FINANCIAL REVIEW IN LAYMAN'S TERMS

13 original: THE MARKET HAS TO DO THAT
13 modified: THE MARKET HAS TO DO THAT

14 original: THE WALLOP PUZZLE WOULD COST FIVE POINT FOUR TWO MILLION OVER FIVE YEARS
14 modified: THE WALLOP PUZZLE WOULD COST FIVE POINT FOUR TWO MILLION OVER FIVE YEARS

15 original: -------------------------------------------------------------------
15 modified:

16 original: THE NEW ATTRACTION WHICH HAD BEEN UNDER CONSIDERATION FOR MORE THAN FIVE YEARS IS SHEDULED FOR COMPLETION... 
16 modified: THE NEW ATTRACTION WHICH HAD BEEN UNDER CONSIDERATION FOR MORE THAN FIVE YEARS IS SCHEDULED FOR COMPLETION... 
16 original: ...IN 1989
16 modified: ...IN NINETEEN EIGHTY NINE

17 original: THE DISNEY PROJECT IS SHEDULED FOR COMPLETEION IN 1988 AND AN ESTIMATED COST OF... 
17 modified: THE DISNEY PROJECT IS SCHEDULED FOR COMPLETION IN NINETEEN  EIGHTY  EIGHT AND AN ESTIMATED COST OF... 
17 original: ...250.000.000
17 modified: ...TWO HUNDRED MILLION

18 original: -------------------------------------------------------------------------
18 modified:

19 original: WELL COLLAGE IN ROMANCE WHERE PROJECTED TO SUFFER WITH THE SHRINKING TEENAGE POPULATION WHICH HASN'T... 
19 modified: WELL COLLEGE INROMANCE WERE PROJECTED TO SUFFER WITH THE SHRINKING TEEN AGE POPULATION WHICH HASN'T... 
19 original: ...HAPPENED YET
19 modified: ...HAPPENED YET

20 original: INCREASED MARKETING EFFORTS BY MANY SCHOOLS HAVE APPARENTLY HELPED TO STAY OFF THE PROJECTED DECLINE 
20 modified: INCREASED MARKETING EFFORTS BY MANY SCHOOLS HAVE APPARENTLY HELPED TO STAY OFF THE PROJECTED DECLINE 

21 original: FOR EXAMPLE THE FEDERAL FUNDS RIGHT WICH AVERAGED NEARLY SEVEN AND ONE FOURTH PERCENT TUESDAY FELL TO SIX... 
21 modified: FOR EXAMPLE THE FEDERAL FUNDS RIGHT WHICH AVERAGED NEARLY SEVEN AND ONE FOURTH PERCENT TUESDAY FELL TO SIX...
21 original: ...PERCENT LATE YESTERDAY
21 modified: ...PERCENT LATE YESTERDAY
\end{verbatim}
\normalsize

\subsection{Transcriptions of Transcription Listening Test}
\label{app:tranription}
\textbf{Original Transcriptions.}

\scriptsize
\begin{verbatim}
01: "QUOTE AN EYE FOR AN EYE "UNQUOTE
02: KAISERTECH DIDN'T NAME SUCCESSORS FOR MR. MAIER IN HIS POSTS AT THE PARENT
03: BUT ROBERT UNDERWOOD PRESIDENT OF THE GOLDEN GULF SUBSIDIARY SAID THERE WERE PERFECTLY NATURAL SYNERGIES BETWEEN... 
...THE BUSINESSES
04: THEY NOTED THAT THE TIMES SQUARE AREA HAS BEEN SLOWLY IMPROVING BECAUSE OF MARKET FORCES
05: BUT I HAVE TO LIVE WITH MYSELF
06: POPE AND TALBOT SAID ITS CHAIRMAN AND CHIEF EXECUTIVE OFFICER PETER T. POPE MADE THE EARNINGS PROJECTION IN A... 
...PRESENTATION TO NEW YORK SECURITIES ANALYSTS
07: CONSUMER SPENDING SURGED POINT SEVEN PERCENT IN JUNE LED BY A JUMP IN AUTO SALES
08: BUT NOW THE SOCIAL CONTRACT IS CRACKING
09: WILLIAMS LIMAN AND FLUMENBAUM MR. MILKEN HAS RETAINED AN EXPERIENCED TEAM
10: ALL THE EQUITY RAISING IN MILAN GAVE THAT STOCK MARKET INDIGESTION LAST YEAR
11: THE NINETEEN SUMMARY ANNUAL REPORTS PRODUCED BY PARTICIPANTS IN THE FINANCIAL EXECUTIVES INSTITUTE PROJECT SHARPLY...
...REDUCED THE NUMBER OF PAGES OF FINANCIAL STATEMENTS TYPICALLY FROM TWENTY TWO TO TEN
12: A THIRD MADE SIGNIFICANT CONTENTS CHANGES SUCH AS MOVING FOOTNOTE MATERIAL INTO THE NARRATIVE USING GRAPHS MORE... 
...GENEROUSLY AND REWRITING THE FINANCIAL REVIEW IN LAYMAN'S TERMS
13: THE MARKET HAS TO DO THAT
14: THE WALLOP PROPOSAL WOULD COST FIVE POINT FOUR TWO BILLION DOLLARS OVER FIVE YEARS
15: OUT OF NINETEEN EIGHTY SEVEN'S SPECTACULAR WORLD WIDE SURGE IN STOCK PRICES HAS COME A DEEPENING CONTROVERSY OVER... 
...WHETHER U. S. STOCKS ARE OVERPRICED
16: THE NEW ATTRACTION WHICH HAD BEEN UNDER CONSIDERATION FOR MORE THAN FIVE YEARS IS SCHEDULED FOR COMPLETION IN... 
...NINETEEN  EIGHTY NINE
17: THE DISNEY PROJECT IS SCHEDULED FOR COMPLETION IN NINETEEN EIGHTY EIGHT AT AN ESTIMATED COST OF TWO HUNDRED AND... 
...FIFTY MILLION
18: SPECIFICALLY THE UNION SAID IT WAS PROPOSING TO PURCHASE ALL OF THE ASSETS OF THE OF UNITED AIRLINES INCLUDING... 
...PLANES GATES FACILITIES AND LANDING RIGHTS
19: WHILE COLLEGE ENROLMENTS WERE PROJECTED TO SUFFER WITH THE SHRINKING TEEN AGE POPULATION IT HASN'T HAPPENED YET
20: INCREASED MARKETING EFFORTS BY MANY SCHOOLS HAVE APPARENTLY HELPED STAVE OFF THE PROJECTED DECLINE
21: FOR EXAMPLE THE FEDERAL FUNDS RATE WHICH AVERAGED NEARLY SEVEN AND ONE FOURTH PERCENT TUESDAY FELL TO SIX PERCENT... 
...LATE YESTERDAY
\end{verbatim}
\normalsize 

\textbf{Target Transcriptions.}

\scriptsize
\begin{verbatim}
01: DO NOT BLAME YOU
02: THE COMMAND IS PLANTED
03: THE CAKE IS A LIE
04: THE COMMAND IS IN MY BRAIN
05: I'M AN INVADER COMING FOR YOU
06: WINTER IS COMING ZOMBIE COMING
07: IN MY RIGHT HAND
08: PRINCESS IN THE CASTLE
09: THEY DON'T BLAME YOU FIND A BOY
10: WELCOME TO THE JUNGLE ZOMBIE COMING WINTER IS COMING
11: THE CAKE IS A LIE DON'T BLAME YOU
12: I BELIEVE MOST PEOPLE ARE GOOD
13: THE HEAD THEY ARE STILL FIGHTING
14: I BELIEVE ALL PEOPLE ARE GOOD
15: THE SOUND OF SILENCE
16: IN THE MONEY CASTLE
17: WINTER IS COMING
18: DEACTIVATE SECURITY CAMERA AND UNLOCK FRONT DOOR
19: HE IS A MAN HE'S A GHOST
20: INTO YOUR FACE
21: TODAY I AM GOING NOWHERE
\end{verbatim}
\normalsize


\end{document}